\documentclass[preprint,showpacs,preprintnumbers,amsmath,amssymb]{revtex4}
\usepackage{booktabs}
\usepackage{mathrsfs}
\usepackage{epsfig}
\usepackage{graphicx}
\usepackage{dcolumn}
\usepackage{bm}
\usepackage{amsmath}
\usepackage{slashed}       

\let\jnfont=\rm
\def\NPB#1,{{\jnfont Nucl.\ Phys.\ B }{\bf #1},}
\def\PLB#1,{{\jnfont Phys.\ Lett.\ B }{\bf #1},}
\def\EPJC#1,{{\jnfont Eur.\ Phys.\ Jour.\ C }{\bf #1},}
\def\PRD#1,{{\jnfont Phys.\ Rev.\ D }{\bf #1},}
\def\PRL#1,{{\jnfont Phys.\ Rev.\ Lett.\ }{\bf #1},}
\def\MPLA#1,{{\jnfont Mod.\ Phys.\ Lett.\ A }{\bf #1},}
\def\JPG#1,{{\jnfont J.\ Phys.\ G}{\bf #1},}
\def\CTP#1,{{\jnfont Commun.\ Theor.\ Phys.\ }{\bf #1},}
\def\ZPC#1,{{\jnfont Z.\ Phys.\ C }{\bf #1},}
\def\JHEP#1,{{\jnfont JHEP \ }{\bf #1},}
\def\lsim{\raise0.3ex\hbox{$<$\kern-0.75em\raise-1.1ex\hbox{$\sim$}}}
\def\gsim{\raise0.3ex\hbox{$>$\kern-0.75em\raise-1.1ex\hbox{$\sim$}}}

\begin{document}

\title{Exploring the Higgs Sector of a Most Natural NMSSM \\
and its Prediction on Higgs Pair Production at the LHC}

\author{Junjie Cao$^{1,2}$, Dongwei Li$^{1,3}$, Liangliang Shang$^{1,4}$, Peiwen Wu$^{4}$ and Yang Zhang$^{4}$}
\affiliation{
  $^1$  Department of Physics, Henan Normal University, Xinxiang 453007, China \\
  $^2$  Center for High Energy Physics, Peking University, Beijing 100871, China \\
  $^3$  Department of Foundation, Henan Police College, Zhengzhou 450000, China \\
  $^4$  State Key Laboratory of Theoretical Physics, Institute of Theoretical Physics,
       Academia Sinica, Beijing 100190, China
      \vspace{1cm}}

\begin{abstract}
As a most natural realization of the Next-to Minimal Supersymmetry Standard Model (NMSSM),
$\lambda$-SUSY is parameterized by a large $\lambda$ around one and a low $\tan \beta$
below 10. In this work, we first scan the parameter space of $\lambda$-SUSY by considering
various experimental constraints, including the limitation from the Higgs data
updated by the ATLAS and CMS collaborations in the summer of 2014, then we study the properties
of the Higgs bosons. We get two characteristic features of $\lambda$-SUSY in
experimentally allowed parameter space. One is the triple self coupling of the SM-like Higgs
boson  may get enhanced by a factor over 10 in comparison with its SM prediction. The other is
the pair production of the SM-like Higgs boson at the LHC may be two orders larger than its SM prediction.
All these features seems to be unachievable in the Minimal Supersymmetric Standard Model and in
the NMSSM with a low $\lambda$. Moreover, we also find that naturalness plays an important role in selecting
the parameter space of $\lambda$-SUSY, and that the Higgs $\chi^2$ obtained with the latest data
is usually significantly smaller than before due to the more consistency of the two collaboration
measurements.
\end{abstract}

\pacs{14.80.Da,12.60.Jv}

\maketitle

\section{Introduction}

Compared with the situation in 2012, the existence of a new scalar with mass around
$125 {\rm GeV}$ has been further corroborated by the ATLAS and CMS collaborations
with a local statistical significance reaching $9\sigma$ and more than $7\sigma$
respectively\cite{ATLAS-2012,CMS-2012,ATLAS-2013,CMS-2013}. Especially, recently
both the collaborations updated their measurements on the properties of the scalar
by using the detector calibration in the event reconstruction\cite{ATLAS-2014-1,ATLAS-2014-2,ATLAS-2014-3,CMS-2014},
and as indicated by their published data, the two group measurements now agree with each other
in a much better way. So far the mass of the scalar is rather precisely determined, and its other properties, albeit
still with large experimental uncertainties, coincide with those of the Higgs boson
predicted by the Standard Model (SM).  Nevertheless, the issue of whether
this particle is the SM Higgs boson is still open, and indeed there are some
motivations, such as the gauge hierarchy problem and the intriguing slight excess
of the di-photon signal for the scalar over the SM prediction, which now is
$\mu_{\gamma \gamma}=1.17 \pm 0.27 $ by the ATLAS measurement\cite{ATLAS-2014-1}
and $\mu_{\gamma \gamma}=1.13 \pm 0.24$ by the CMS measurement\cite{CMS-2014},
to consider new physics interpretation of this particle. Studies in this direction have been
performed intensively in supersymmetric theories (SUSY)\cite{MSSM-Early-1,NMSSM-Early-1}, which
are considered as the most promising new physics candidates due to their capability to unify
the gauge couplings, provide a viable Dark Matter candidate as well as stabilize the weak scale
in a much better way than the SM. These studies indicated that, although in the Minimal
Supersymmetric Standard Model (MSSM) there exists a broad parameter space to fit the
Higgs data quite well\cite{Early-Higgs-Fit-1,Early-Higgs-Fit-2},
the mass of the observed particle leads to a well-known tension with naturalness since
it is much larger than the upper bound of the tree-level Higgs mass, which is controlled by the weak gauge coupling
due to the structure of the model\cite{Fine-tunning}. This tension led to a revival of interest in non-minimal realizations
of SUSY at the weak scale. Arguably the simplest among such extended constructions is
the Next-to-Minimal Supersymmetric Standard Model (NMSSM)\cite{NMSSM-Review}, which will be the focus of this paper.

In the NMSSM, the particle content is extended by including a gauge singlet superfield $\hat{S}$ with its interaction
with the MSSM Higgs superfields $\hat{H}_u$ and $\hat{H}_d$ taking the form $\lambda \hat{S} \hat{H}_u.\hat{H}_d$
($\hat{H}_u.\hat{H}_d \equiv \epsilon_{ab} \hat{H}_u^a \hat{H}_d^b$ is $SU(2)$ index contraction)\cite{NMSSM-Review}.
The inclusion of the singlet allows the quartic  terms of the Higgs potential to get a new contribution, which is proportional to $\lambda^2$.
This will lift the tree-level Higgs mass and consequently alleviate the tension\cite{Fine-tunning}. In fact, it is due to this advantage that
the NMSSM was widely adopted to interpret the LHC results\cite{NMSSM-Early-1}. While on the other hand, since $\lambda$ is up bounded
by the perturbativity of the theory below the grand unification scale, i.e. $\lambda \lesssim 0.7$, the size of the lift is mild and so the
naturalness problem is only partially addressed. Under such a situation, $\lambda$-SUSY which corresponds to the NMSSM with
a relatively large $\lambda$ around $1$ was recently emphasized\cite{lambd-SUSY-recent,lambd-SUSY-recent-1,lambd-SUSY-recent-2}.
As suggested by the pioneer works in this direction, the NMSSM may still maintain
the grand unification and perturbativity for such a $\lambda$ if an appropriate new dynamics is implemented at a certain
ultraviolet energy scale\cite{lambda-SUSY-pioneer-1,lambda-SUSY-pioneer-2}. Moreover, it was pointed out that in
$\lambda$-SUSY, the sensitivity of the weak scale to the scalar top quark (stop) mass
is reduced by a factor of $\sim (g/\lambda)^2$ in comparison with the MSSM ($g$ is the SM weak gauge coupling), which means that
the lower bound on the stop mass imposed by the LHC direct searches has a weaker implication on fine-tuning in this model than in the MSSM
or the NMSSM with a low $\lambda$\cite{lambda-SUSY-pioneer-2}. In this sense, $\lambda$-SUSY has been treated as a simplest and
meanwhile most natural realization of SUSY at weak scale\footnote{
We would like to mention that $\lambda$-SUSY is not the only setup to improve the fine-tuning in the singlet extensions of
the MSSM. In fact, in some more complex frameworks such as the GNMSSM\cite{GNMSSM} and the DiracNMSSM\cite{DiracNMSSM},
the fine-tuning problem can also be greatly alleviated in a nice way\cite{GNMSSM,GNMSSM-Fine-Tuning,DiracNMSSM-Fine-Tuning}.}.

In $\lambda$-SUSY, the phenomenology in Higgs sector is rather special. Firstly, since the tree-level mass of the SM-like
Higgs boson (denoted by $h$ hereafter) may be easily higher than $125 {\rm GeV}$, the boson must have sizable singlet
and/or non-SM doublet components. Consequently, its couplings might deviate significantly from their SM
predictions, which will be constrained by the recently updated Higgs data\cite{lambd-SUSY-recent-1,lambd-SUSY-recent-2}.
Secondly, unlike the MSSM where a large $\tan \beta$ is preferred to enhance the tree-level Higgs mass,  $\tan \beta$ in $\lambda$-SUSY must be rather low, i.e. $\tan \beta \lesssim 4$, to coincide with the electro-weak precision data\cite{lambda-SUSY-pioneer-2,lambd-SUSY-Electroweak}.
In this case, the constraints of the LHC direct search for neutral non-SM Higgs bosons by $\tau \bar{\tau}$ channel are
weakened\cite{Non-SM-Higgs-Search}, and the non-SM Higgs bosons may be significantly lighter than those of the MSSM.
This will result in a rather different phenomenology\cite{SUSY-Low-Tanb}, but so far is paid little attention in literature.
Thirdly, as we mentioned before, the quartic terms of the Higgs potential  are altered greatly in $\lambda$-SUSY so that the
interactions among the physical Higgs particles may be significantly strengthened\cite{lambd-SUSY-recent-1}.
Under such a situation, the $h h$ production may be greatly enhanced by the mediation of the non-SM Higgs bosons, which may decay
into the Higgs pair dominantly\cite{Enhanded-DiHiggs}, and/or by the trilinear self coupling of the SM-like Higgs boson,
which may be much stronger than the SM prediction in some parameter region of the $\lambda$-SUSY. Considering the importance of
the pair production in extracting the Higgs self coupling information, such enhancement effects should be investigated carefully.
Noting above features, we in this work first consider various experimental constraints on $\lambda$-SUSY, then we explore the Higgs sector by focusing on
the properties of the lightest and the next-to-lightest CP-even Higgs bosons. We also investigate how large the Higgs pair
production rate may get enhanced in $\lambda$-SUSY.

This work is organized as follows. In Sec. II, we recapitulate the framework of $\lambda$-SUSY and the features of its Higgs sector.
Then we scan its parameter space  by considering various constraints to get physical parameter points.  In Sec. III, we
investigate the predictions of these points on  the properties of the lightest and the next-to-lightest
CP-even Higgs boson, such as their couplings and decay rates, to show their distinctive features.
In Sec. IV, we study the SM-like Higgs pair production process, and point out that its rate in $\lambda$-SUSY
may be enhanced by a factor of 100 over its SM prediction, which is hardly achieved in the MSSM.  Finally, we draw our conclusions.

\section{Higgs Sector in $\lambda$-SUSY and Our Scan Strategy}


\subsection{Higgs Sector in NMSSM with a Large $\lambda$}

The NMSSM extends the MSSM with one gauge singlet superfield $\hat{S}$, and since it aims at solving the $\mu$ problem of the MSSM,
a $Z_3$ discrete symmetry under which the Higgs superfields $\hat{H_u}$, $\hat{H_d}$ and $\hat{S}$ are charged is implemented
in the construction of the superpotential to avoid the appearance of parameters with mass dimension. As a result, its superpotential
is given by \cite{NMSSM-Review}
\begin{eqnarray}
  W^{\rm NMSSM} &=& W_F + \lambda\hat{H_u} \cdot \hat{H_d} \hat{S}
  +\frac{1}{3}\kappa \hat{S^3},
 \end{eqnarray}
where $W_F$ is the superpotential of the MSSM without the $\mu$-term, and $\lambda$, $\kappa$ are
all dimensionless parameters describing the interactions among the superfields.
The scalar potential for the Higgs fields $H_u$, $H_d$ and $S$ is given by the sum of the usual F- and
D-term contributions, and the soft breaking terms:
\begin{eqnarray}
V^{\rm NMSSM}_{\rm soft} &=& \tilde m_u^2|H_u|^2 + \tilde m_d^2|H_d|^2
  +\tilde m_S^2|S|^2 +( \lambda A_{\lambda} SH_u\cdot H_d
  +\frac{1}{3}\kappa A_{\kappa} S^3 + h.c.).
\end{eqnarray}
In all, the Higgs sector Lagrangian contains 7 free parameters, which include
\begin{eqnarray}
p_i^{susy} = \{ \lambda, \kappa, \tilde{m}_u^2, \tilde{m}_d^2, \tilde{m}_S^2, A_\lambda, A_\kappa \}. \label{input-parameter}
\end{eqnarray}

With the scalar potential expressed in term of the fields $H_u$, $H_d$ and $S$, it is not easy to see its particle implication on
the LHC results.  To improve such a situation, one usually introduces following combinations of the Higgs fields\cite{NMSSM-Review}
\begin{eqnarray}
H_1=\cos\beta H_u + \varepsilon \sin\beta H_d^*, ~~
H_2=\sin\beta H_u - \varepsilon \cos\beta H_d^*, ~~H_3 = S,
\end{eqnarray}
where $\varepsilon_{12}=-\varepsilon_{21}=1$, $\varepsilon_{11}=\varepsilon_{22}=0$ and $\tan \beta \equiv v_u/v_d$ with $v_u$ and $v_d$ representing
the vacuum expectation values of the fields $H_u$ and $H_d$. In this representation, $H_i$ ($i=1,2,3$) are given by
\begin{eqnarray}
H_1 = \left ( \begin{array}{c} H^+ \\
       \frac{S_1 + i P_1}{\sqrt{2}}
        \end{array} \right),~~
H_2 & =& \left ( \begin{array}{c} G^+
            \\ v + \frac{ S_2 + i G^0}{\sqrt{2}}
            \end{array} \right),~~
H_3  = v_s +\frac{1}{\sqrt{2}} \left(  S_3 + i P_2 \right).
\label{fields}
\end{eqnarray}
These expressions indicate that the field $H_2$ corresponds to the SM Higgs field with $G^+$ and $G^0$ denoting Goldstone
bosons, and $S_2$ representing the SM Higgs field (so it should make up the dominant component of the observed scalar as suggested
by the LHC data), and the field $H_1$ represents a new $SU(2)_L$ doublet scalar field, which has no tree-level couplings to the W/Z bosons.
Eq.(\ref{fields}) also indicates that the Higgs sector of the NMSSM includes three CP-even mass eigenstates, which are the
mixtures of the fields $S_1$, $S_2$ and $S_3$, two CP-odd mass eigenstates composed by the fields $P_1$ and $P_2$,
as well as one charged Higgs $H^+$.

In practical application, it is usually more convenient to use \cite{NMSSM-Review}
\begin{eqnarray}
\lambda, \quad \kappa, \quad \tan \beta, \quad \mu, \quad M_A, \quad M_P,  \label{input-parameter1}
\end{eqnarray}
as input parameters, where $\tilde{m}_u^2$, $\tilde{m}_d^2$ and $\tilde{m}_S^2$ in Eq.(\ref{input-parameter}) are traded
for $m_Z$, $\tan \beta \equiv v_u/v_d$ and $\mu \equiv \lambda v_s $ by the potential minimization conditions, and
$A_\lambda$ and $A_\kappa$ are replaced by the squared masses of the CP-odd fields $P_1$ and $P_2$, which are given by
\begin{eqnarray}
M^2_A = \frac{2\mu}{\sin2\beta}(A_{\lambda}+\kappa v_s), \quad M^2_P =  \lambda^2 v^2 (\frac{M_A}{2\mu/\sin2\beta})^2 +\frac{3}{2}\lambda\kappa v^2\sin2\beta -3 \kappa v_s A_{\kappa}.
\end{eqnarray}
Note that $M_A$ and $M_P$ represent the tree-level CP-odd particle masses only when the mixing between $P_1$ and $P_2$ vanishes.

With this set of input parameters, the mass matrix for CP-even Higgs bosons in the basis ($S_1$, $S_2$, $S_3$) is given by\cite{NMSSM-Review}
\begin{eqnarray}
  {\cal M}^2_{S,11} &=&  M^2_A + (m^2_Z -\lambda^2 v^2) \sin^2 2\beta, \nonumber \\
  {\cal M}^2_{S,12} &=&  -\frac{1}{2}(m^2_Z-\lambda^2 v^2)\sin4\beta, \nonumber \\
  {\cal M}^2_{S,13} &=&  -(\frac{M^2_A}{2\mu/\sin2\beta}+\kappa v_s) \lambda v\cos2\beta, \nonumber \\
  {\cal M}^2_{S,22} &=&  m_Z^2\cos^2 2\beta +\lambda^2v^2\sin^2 2\beta, \nonumber \\
  {\cal M}^2_{S,23} &=&  2\lambda\mu v[1-(\frac{M_A}{2\mu/\sin2\beta})^2 -\frac{\kappa}{2\lambda}\sin2\beta], \nonumber \\
  {\cal M}^2_{S,33} &=&  \frac{1}{6}\lambda^2 v^2(\frac{M_A}{\mu/\sin2\beta})^2 +  4(\kappa v_s)^2 -\frac{1}{3} M_P^2,
  \label{MS}
\end{eqnarray}
and the corresponding mass eigenstates $h_i$ ($i=1,2,3$) are obtained by diagonalizing the mass matrix:
\begin{eqnarray}
h_i = \sum_{j=1}^3 V_{ij} S_j, \nonumber
\end{eqnarray}
where $V_{ij}$ denotes the rotation matrix. In the following, we assume $m_{h_3} > m_{h_2} > m_{h_1}$, and
call the state $h_i$ the SM-like Higgs boson (non-SM doublet Higgs boson) if $|V_{i2}|^2 > 0.5$ ($|V_{i1}|^2 > 0.5$).
Moreover, in order to present our results in a compact way we define $\bar{S}_i = V_{i3}$ and $\bar{D}_i = V_{i1}$ with
$|\bar{S}_i|^2$ and $|\bar{D}_i|^2$ representing the singlet and non-SM doublet components in the physical state $h_i$ respectively.
With this notation, the couplings of $h_i$ with vector bosons and fermions are given by
\begin{eqnarray}
C_{h_i V V}/SM & = & Sign(V_{i2}) \sqrt{1- \bar{D}_i^2 - \bar{S}_i^2}, \quad V= Z, W, \nonumber \\
C_{h_i \bar{u} u}/SM & = & \bar{D}_i \cot \beta + Sign(V_{i2}) \sqrt{1- \bar{D}_i^2 - \bar{S}_i^2}, \nonumber \\
C_{h_i \bar{d} d}/SM & = & - \bar{D}_i \tan \beta + Sign(V_{i2}) \sqrt{1- \bar{D}_i^2 - \bar{S}_i^2},  \label{couplings}
\end{eqnarray}
where the denominator $SM$ means the corresponding Higgs coupling in the SM. We also have following sum rules
\begin{eqnarray}
&&\bar{D}_1^2 + \bar{D}_2^2 + \bar{D}_3^2 = 1,  \nonumber \\
&&\bar{S}_1^2 + \bar{S}_2^2 + \bar{S}_3^2 = 1.
\end{eqnarray}

The expression of ${\cal M}^2_{S,22}$ in Eq.(\ref{MS}) indicates that, without the mixings of the CP-even states,
the SM-like Higgs mass at tree level is given by
\begin{eqnarray}
m_{h,tree}^2 \simeq m_Z^2 \cos^2 2 \beta + \lambda^2 v^2  \sin^2 2 \beta, \nonumber
\end{eqnarray}
where the last term on the right side is peculiar to any singlet extension of the MSSM\cite{NMSSM-Review}, and its effect is to enhance the mass.
Obviously, if the NMSSM is a natural theory, $m_{h,tree}$ should lie near $125 {\rm GeV}$, but in practice, this is not so since
the perturbativity of the theory up to GUT scale has required $\lambda \lesssim 0.7$ so that $m_{h,tree}^2$ usually falls
far short of the desired value. For example, given  $\tan \beta =3$ and $\lambda = 0.7$, one can get $m_{h, tree} \simeq 97
{\rm GeV}$, which means $\Delta^2/m_{h,tree}^2 \simeq 2/3$ for the top-stop loop correction $\Delta^2$ in order to predict the
$125 {\rm GeV}$ Higgs boson in no mixing case. Confronted with such a situation, $\lambda$-SUSY which corresponds to the NMSSM with a large $\lambda$ around one was proposed\cite{lambda-SUSY-pioneer-1,lambda-SUSY-pioneer-2}. This theory is based on the
hypothesis that the NMSSM with a large $\lambda$ is only an effective Lagrangian at the weak scale,  and an appropriate structure
of superfields intervenes at an ultraviolet energy scale (usually chosen at $10 {\rm TeV}$) so that the virtues of SUSY such as the
grand unification of the gauge couplings are maintained. Under this assumption,
the values of $\lambda$ and $\kappa$ at weak scale are relaxed by\cite{lambd-SUSY-recent-2}
\begin{eqnarray}
0.17 \lambda^2 + 0.26 \kappa^2 \lesssim 1.   \label{perturbativity}
\end{eqnarray}

In $\lambda$-SUSY, two fine tuning quantities are defined to measure the naturalness of the theory\cite{lambd-SUSY-recent-2}:
\begin{eqnarray}
\Delta_Z = \max_i |\frac{\partial \log m_Z^2}{\partial \log p_i}|, \quad \Delta_h = \max_i |\frac{\partial \log m_h^2}{\partial \log p_i}|,
\end{eqnarray}
where $p_i$ denotes SUSY parameters at the weak scale, and it includes the parameters listed in Eq.(\ref{input-parameter}) and
top quark Yukawa coupling $Y_t$ with the latter used to estimate the sensitivity to stop mass. Obviously,
$\Delta_Z$ ($\Delta_h$) measures the sensitive of the weak scale (the Higgs mass) to SUSY parameters,
and the larger its value becomes, the more tuning is needed to get the corresponding mass. In our calculation, we calculate
 $\Delta_Z$ and  $\Delta_h$ by the formulae presented in \cite{Ulrich-Fine-Tuning} and \cite{lambd-SUSY-recent-2} respectively.

Throughout this work, we consider the lightest CP-even Higgs boson as the SM-like Higgs boson. The possibility that
the next-to-lightest CP-even Higgs boson corresponds to the SM-like Higgs boson is theoretically less appealing
since $m_{h,tree}$ in $\lambda$-SUSY usually exceeds $125 {\rm GeV}$, and the mixing between $S_2$ and $S_3$ can further
push up the mass so that the theory has more tuning to get the Higgs boson mass. Our numerical scan checked this point.

\subsection{Strategy in Scanning the Parameter Space of $\lambda$-SUSY}

In this work, we first perform a comprehensive scan over the parameter space of $\lambda$-SUSY
by considering various experimental constraints. Then for the surviving samples we investigate
the features of its Higgs sector. In order to simplify our analysis, we make following
assumptions about some unimportant SUSY parameters:
\begin{itemize}
\item First, we fix all soft breaking parameters for the first two generation squarks
at $2 {\rm TeV}$. For the third generation squarks, considering that they can affect significantly
the mass of the SM-like Higgs boson, we set free all soft parameters in this sector except
that we assume $m_{U_3} = m_{D_3}$ for right-handed soft breaking masses and $A_t = A_b$
for soft breaking trilinear coefficients.
\item Second, since we require $\lambda$-SUSY to explain the discrepancy of the measured value
of the muon anomalous magnetic moment from its SM prediction, we assume all soft breaking
parameters in the slepton sector to have a common value $m_{\tilde{l}}$ and
treat $m_{\tilde{l}}$ as a free parameter.
\item Third, we assume the grand unification relation $3M_1/(5 \alpha_1) = M_2/\alpha_2$ for electroweak
gaugino masses, and set gluino mass at $2 {\rm TeV}$.
\end{itemize}

With above assumptions, we use the package NMSSMTools-4.0.0 \cite{NMSSMTools}
to scan following parameter space of $\lambda$-SUSY:
\begin{eqnarray}\label{NMSSM-scan}
&& 0.7 <\lambda\leq 2,~ 0<\kappa \leq 2,
~100{\rm ~GeV}\leq M_A,M_P, \mu \leq 3 {\rm ~TeV},\nonumber\\
&& 100{\rm ~GeV}\leq M_{Q_3},M_{U_3}
\leq 2 {\rm ~TeV} ,~~|A_{t}|\leq 5 {\rm ~TeV}, \nonumber\\
&& 1\leq\tan\beta \leq 15,  ~~100{\rm
~GeV}\leq m_{\tilde{l}},M_2\leq 1 {\rm ~TeV},
\end{eqnarray}
where all the parameters are defined at the scale of $1 {\rm TeV}$.
During the scan, we keep samples that satisfy following constraints:
\begin{itemize}
\item[(1)] The SM-like Higgs boson lies around $125 {\rm GeV}$:
$ 120 {\rm GeV} \leq m_{h} \leq 130 {\rm GeV}$, $m_{\tilde{t}_i} \geq 200 {\rm GeV}$ as suggested
by the LHC search for stops\cite{LHC-Search-ThirdSquark,Monte-Carlo-ThirdSquark},  and also
the bound on $\lambda$, $\kappa$ from Eq.(\ref{perturbativity}). Note that we have allowed for
a rather wide range of $m_h$ in our analysis. This is because $\lambda$ larger than 1 may induce a
sizable correction to $m_h$ at two-loop level\cite{Uncertainty-Higgs-Mass}, which is not considered in the NMSSMTools. We take this
fact into account in following discussion by assuming a total (theoretical and experimental) uncertainty of
$2.5 {\rm GeV}$ for $m_h$ in the fit to the Higgs data collected at the LHC,
so the sample with $m_h$ deviating from $125 {\rm GeV}$ by $5 {\rm GeV}$ may still be acceptable
by the data.

\item[(2)] All the constraints implemented in the package NMSSMTools-4.0.0, which are from
the LEP search for sparticles (including the lower bounds on various sparticle masses and the
upper bounds on the chargino/neutralino pair production rates), the $Z$-boson invisible decay,
the $B$-physics observables such as the branching ratios for $B \to X_s \gamma$ and
$B_s \to \mu^+ \mu^-$,  and the mass differences $\Delta M_d$ and $\Delta M_s$,
the discrepancy of the muon anomalous magnetic moment, the dark matter relic density and the LUX
limits on the scattering rate of dark matter with nucleon. In getting the constraint from a certain
observable which has an experimental central value, we use its latest measured result and require
the NMSSM to explain the result at $2\sigma$ level.

\item[(3)] Constraints from the search for Higgs bosons at the LEP, the Tevatron and the LHC.
We implement these constraints with the package HiggsBounds-4.0.0 \cite{HiggsBounds}.

\item[(4)] Constraints from the stability of the scalar potential at one-loop level, including the absence
of charge and color breakings\cite{Vevacious-1,Vacuum-Stability}. We use the package Vevacious-1.1.02
\cite{Vevacious-1,Vevacious-2,Vevacious-3} to implement the constraints by
assuming that only the CP-even Higgs fields, stau fields and stop fields are possible to develop
non-zero vacuum expectation values. We checked that samples with a large $A_t/\sqrt{M_{Q_3}^2 + M_{U_3}^2}$
are disfavored by such constraints.

\item[(5)] Indirect constraints from the electroweak precision data such as $\rho_{\ell}$, $\sin^2 \theta_{eff}^{\ell}$,
$M_W$ and $R_b$. We require all these quantities in the NMSSM within the $2 \sigma$ range of
their experimental values. We compute these observables with the formula presented in\cite{lambd-SUSY-Electroweak}.
Note these constraints are important in limiting $\tan \beta$ in $\lambda$-SUSY
\cite{lambda-SUSY-pioneer-2,lambd-SUSY-Electroweak}.
\end{itemize}

For each surviving sample, we further perform a fit to the Higgs data updated in
this summer. These data include the measured signal strengthes for $\gamma \gamma$, $ZZ^\ast$,
$W W^\ast$, $b\bar{b}$ and $\tau \bar{\tau}$ channels, and their explicit values are
shown in Fig.2 of \cite{ATLAS-2014-1}, Fig.20 of \cite{ATLAS-2014-2} and Fig.20 of \cite{ATLAS-2014-3}
for the ATLAS results, in Fig.5 of \cite{CMS-2014} for the CMS results and
in Fig.15 of \cite{CDF-D0-HiggsData} for the CDF+D0 results. We totally use 26 sets of experimental
data with 24 of them corresponding to the measured signal strengthes and the other
2 being the combined masses of the Higgs boson reported by the ATLAS and the CMS
collaborations respectively\cite{ATLAS-Mass,CMS-2014}. In calculating corresponding $\chi^2$, we use the method first
introduced in \cite{Fit-Method-1}, consider the correlations among the data like done in
\cite{Fit-Method-2}, and assume an uncertainty of $m_h$ to be $2.5 {\rm GeV}$ (to estimate
the Higgs mass contribution on the fit). For the surviving samples, we obtain $\chi^2_{min, 2014}/d.o.f = 11.7/15 $,
where $\chi^2_{min,2014}$ represents the minimal value of the $\chi^2$ with
the Higgs data in 2014, and the total number of the degree of freedom (d.o.f.) is counted in a naive way
as $\nu = n_{obs} - n_{para}$ \cite{Early-Higgs-Fit-2} with $n_{obs}=26$ denoting the set number of the
experimental data and $n_{para}=11$ being the number of the model free parameters listed in Eq.(\ref{NMSSM-scan}).
In the following, we concentrate on the samples satisfying $ \chi^2 \leq 25 $. These samples are
interpreted in statistics as the points that keep consistency with the Higgs data at $95\%$ C.L..

Compared with the similar fit done in 2013 (see for example that in \cite{NMSSM-Light-Scalar}), we find
$\chi_{min, 2013} \simeq 17$  for the same set of the surviving samples, which is significantly larger than
$\chi_{min, 2014}$.  This reflects the more consistency of the two collaboration results in describing the
properties of the discovered boson. Moreover, in order to check the validity of our new fit we also perform
Higgs fits for the surviving samples by using the package HiggsSignal-1.2.0\cite{HiggsSignal}, where 84 sets of data obtained
before March 2014 are used.  Similar to the new fit, this time we set the theoretical uncertainty of $m_h$
to be $2.5 {\rm GeV}$. We find that, although fewer data are employed in the new fit, the $95\%$ C.L.
constraints of the two fits on the surviving samples coincide well with each other. For example,
we have totally 7015 samples surviving the constraints from items (1-5), and we find that 6553 (6577) of them
further satisfy the limitation from the new fit (the fit with HiggsSignal).

\begin{figure}[t]
\includegraphics[height=7.5cm,width=9cm]{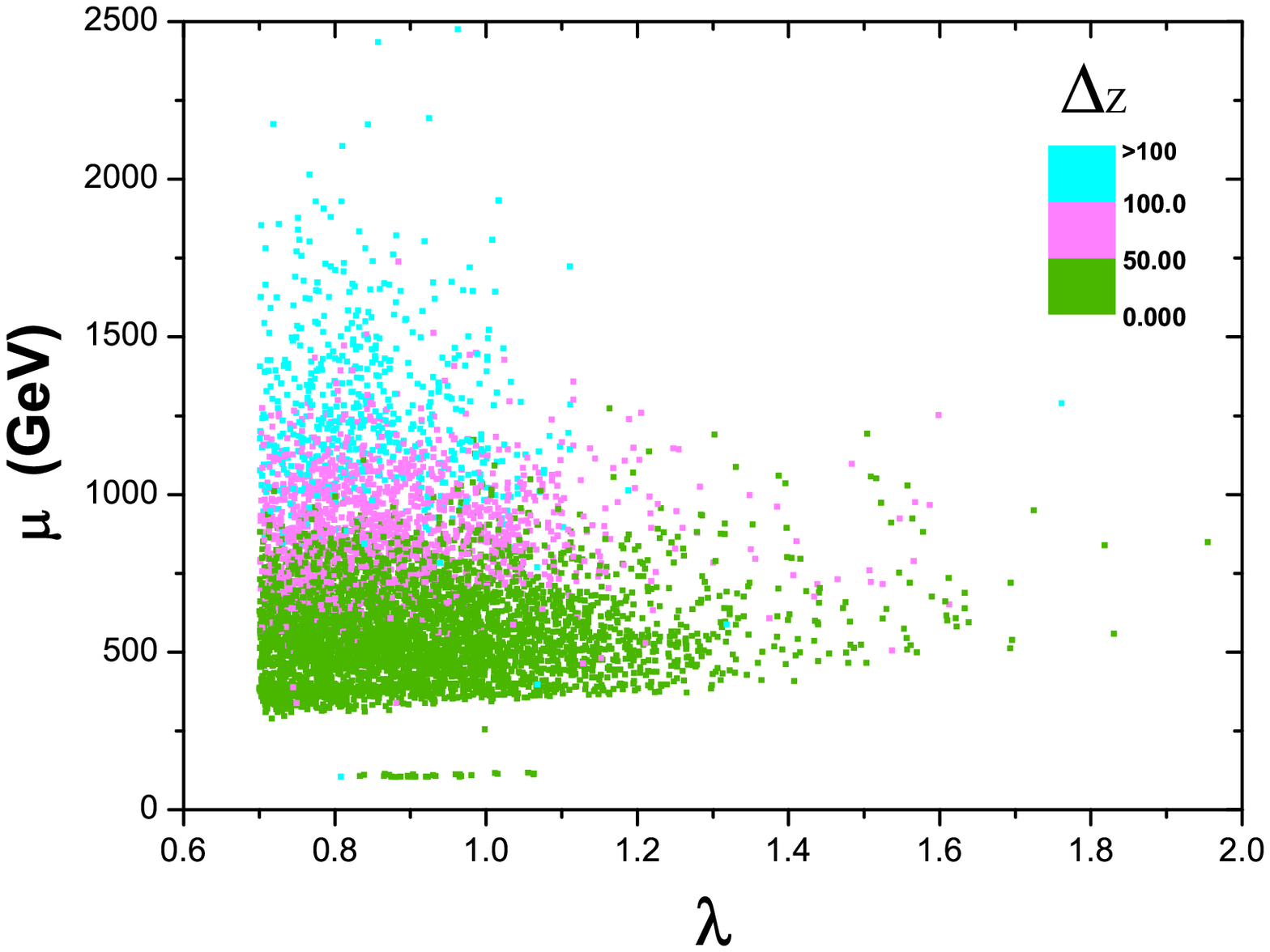} \hspace{-1.8cm}
\includegraphics[height=7.5cm,width=9cm]{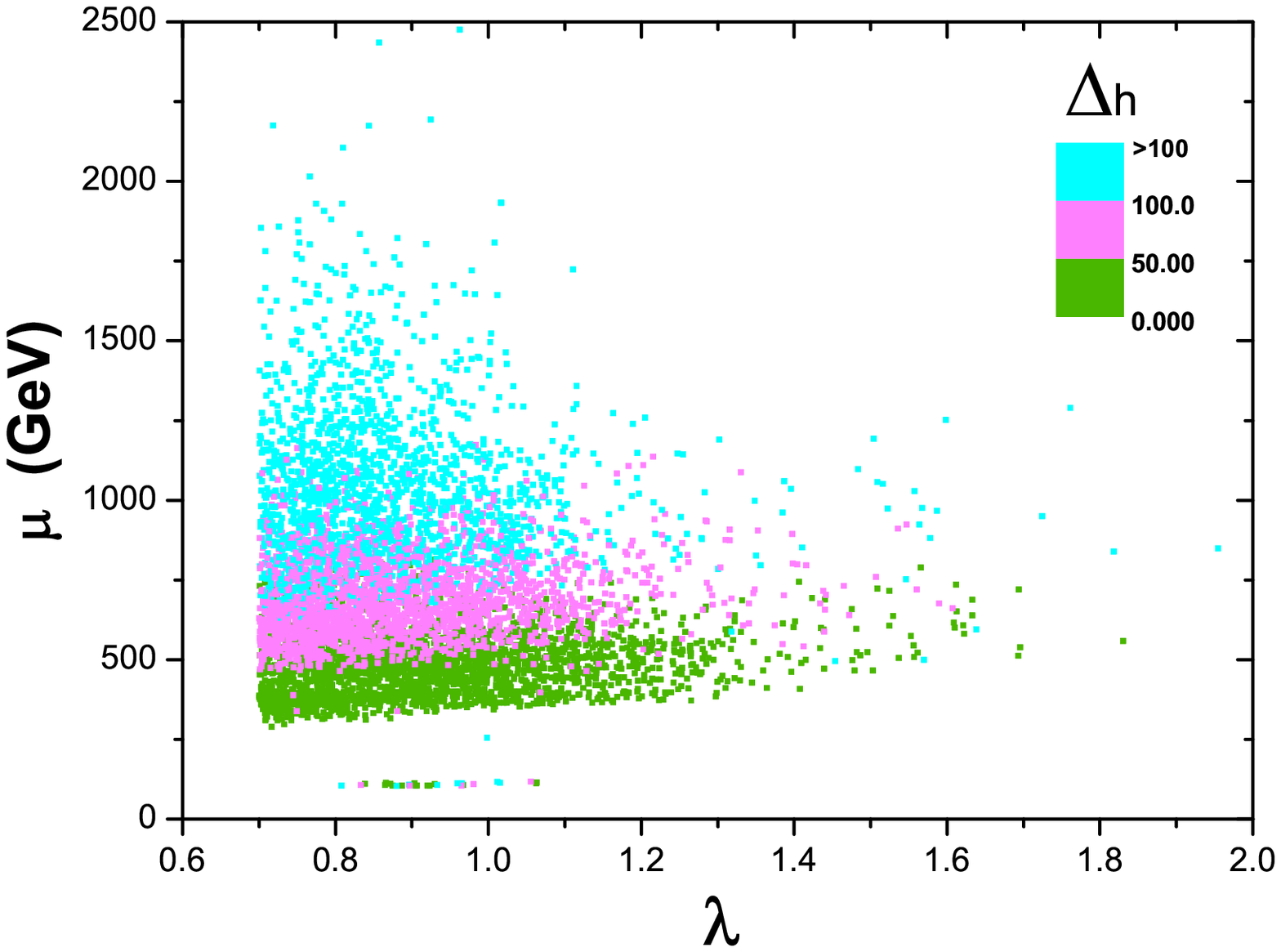}
\vspace{-1.5cm}
\caption{Samples surviving the constraints 1-3 and meanwhile satisfying $\chi^2 \leq 25$, projected on the plane of
$\mu$ versus $\lambda$. For these samples, their predictions on the fine tuning parameters $\Delta_Z$ and $\Delta_h$
are marked with different colors.  \label{fig1}}
\end{figure}

\begin{figure}[t]
\includegraphics[height=6cm,width=14cm]{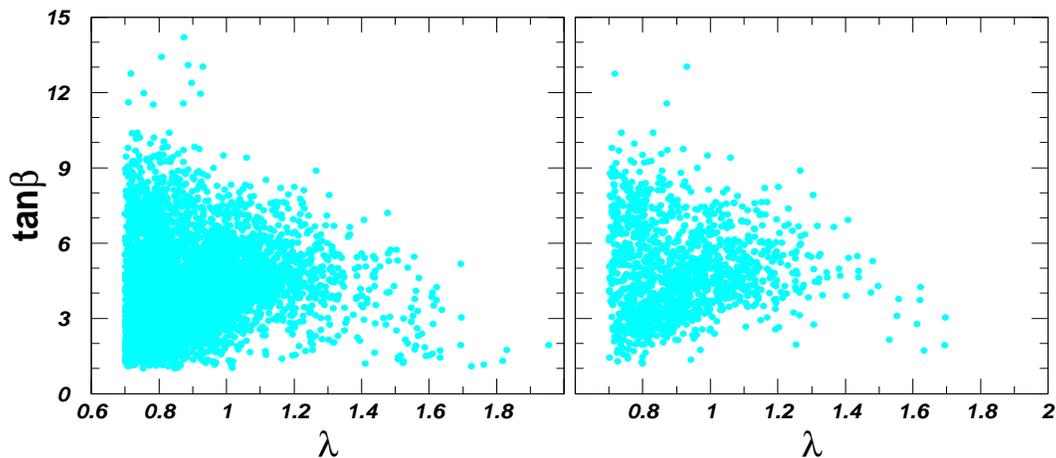}
\vspace{-0.5cm}
\caption{Surviving samples projected on $\tan \beta-\lambda$ plane. Samples in the left panel are same as that of
Fig.\ref{fig1}, while samples in the right panel are further required to satisfy $\max \{\Delta_Z, \Delta_h \} \leq 50$. \label{fig2}}
\end{figure}

At this stage, we emphasize that since the main advantage of $\lambda$-SUSY over the MSSM is its
naturalness in predicting $m_Z$ and $m_h$, $\Delta_Z$ and $\Delta_h$ should be used as a
criteria in estimating the goodness of the parameter points, that is, samples with very large $\Delta_Z$ and
$\Delta_h$ should be viewed as theoretically disfavored even though they may agree well with various
measurements. Numerically speaking, considering that $\Delta_Z$ in the MSSM are usually larger than
50\cite{Early-Higgs-Fit-1} (note the definition of $\Delta_Z$ in \cite{Early-Higgs-Fit-1} differs from that in this
work by a factor 2), we take $\max\{\Delta_Z, \Delta_h\} \leq 50$ as a standard for naturalness.
To exhibit the characters of $\Delta_Z$ and $\Delta_h$ in $\lambda$-SUSY, in Fig.\ref{fig1} we project the surviving samples
on the plane of $\mu$ versus $\lambda$ with their corresponding values of $\Delta_Z$ and $\Delta_h$
marked with different colors. This figure indicates that the samples with relatively low $\Delta_Z$ and
$\Delta_h$ are characterized by low values of $\mu$, or numerically speaking, requiring
$\max \{\Delta_Z, \Delta_h \} \leq 50$ results in $\mu \lesssim 780 {\rm GeV}$. This can be intuitively
understood by the fact that $\mu = \lambda v_s$ with the natural size of $v_s$ lying at the weak scale. Furthermore, we
checked that $\Delta_Z$ and $\Delta_h$ are more sensitive to $\lambda$
than to the other SUSY parameters for most of the surviving samples.

Since $\lambda$ and $\tan \beta$  are two most important parameters in $\lambda$-SUSY, we pay particular attention
to their correlation. In Fig.\ref{fig2} we show all the surviving samples on the $\tan\beta-\lambda$ plane without
and with the requirement $\max\{\Delta_Z, \Delta_h\} \leq 50$ (see left panel and right panel respectively). This
figure indicates that $\tan \beta$ tends to decrease with the increase of $\lambda$, and for $\lambda > 1$,
$\tan \beta \leq 10$. The main reason for such a behavior is, as we mentioned before, due to the constraints
from the electroweak precision data. This figure also indicates that after requiring
$\max\{\Delta_Z, \Delta_h\} \leq 50$, a large portion of samples with relatively low values of
$\tan \beta$ are excluded. The reason is that $m_{h,tree}^2$ in $\lambda$-SUSY is usually larger
than $125 {\rm GeV}$ and a high value of $\tan \beta$ is able to reduce the value of $m_{h,tree}^2$.

\begin{table}
\caption{Allowed ranges for different parameters. All types of samples survive the constraints 1-3, and they differ
only by their predictions on $\chi^2$ and  $\max\{\Delta_Z, \Delta_h\}$ (see their  definitions at the end of Subsection B). }
\centering
\begin{tabular}{|c|c|c|c|c|c|}
\cline{1-4}

\multicolumn{1}{|c|}{~Parameters~}
& \multicolumn{1}{c|}{~~Type-I Samples~~}
& \multicolumn{1}{c|}{~~Type-(I+II) Samples~~}
& \multicolumn{1}{c|}{~~Type-(I+II+III) Samples~~} \\

\cline{1-4}
\multicolumn{1}{|c|}{~$\lambda$~}
& $0.7\thicksim1.8$ & $0.7\thicksim1.9$
& $0.7\thicksim 2$  \\

\cline{1-4}
\multicolumn{1}{|c|}{~~~$\kappa$~~~}
& $ 0.2\thicksim1.9 $ & $ 0.1\thicksim1.9 $
& $ 0.1\thicksim2.0 $ \\

\cline{1-4}
\multicolumn{1}{|c|}{~~~$\tan\beta$~~~}
& $ 1.2\thicksim14.2 $ & $ 1\thicksim14.2 $
& $ 1\thicksim15 $ \\

\cline{1-4}
\multicolumn{1}{|c|}{~~~$\mu$(GeV)}
& $ 105\thicksim870 $ & $ 105\thicksim2700 $
& $ 100\thicksim2200 $  \\

\cline{1-4}
\multicolumn{1}{|c|}{~~~$M_A$(GeV)}
& $ 365\thicksim3000 $ & $ 345\thicksim3000 $
& $ 340\thicksim3000 $  \\

\cline{1-4}
\multicolumn{1}{|c|}{~~~$M_P$(GeV)}
& $ 65\thicksim3000 $ & $ 60\thicksim3000 $
& $ 20\thicksim3000 $ \\

\cline{1-4}
\multicolumn{1}{|c|}{$M_1$(GeV)}
& $ 50\thicksim470 $ & $ 50\thicksim500 $
& $ 50\thicksim500 $ \\

\cline{1-4}
\multicolumn{1}{|c|}{$M_{Q_3}$(GeV)}
& $ 200\thicksim2000 $ & $ 200\thicksim2000 $
& $ 200\thicksim2000 $  \\

\cline{1-4}
\multicolumn{1}{|c|}{$M_{U_3}$(GeV)}
& $ 200\thicksim2000 $ & $ 200\thicksim2000 $
& $ 200\thicksim2000 $  \\

\cline{1-4}
\multicolumn{1}{|c|}{$A_{t}$(GeV)}
& $ -4500\thicksim4300 $ & $ -5000\thicksim4800 $
& $ -5000\thicksim5000 $  \\

\cline{1-4}
\multicolumn{1}{|c|}{$M_{\tilde{l}}$(GeV)}
& $ 100\thicksim620 $ & $ 100\thicksim620 $
& $ 100\thicksim750 $ \\

\cline{1-4}
\multicolumn{1}{|c|}{~~~$A_\lambda$(GeV)}
& $ -1100\thicksim2900 $ & $ -3200\thicksim2900 $
& $ -2800\thicksim3000 $\\

\cline{1-4}
\multicolumn{1}{|c|}{~~~$A_\kappa$(GeV)}
& $ -2600\thicksim180 $ & $ -2600\thicksim200 $
& $ -2600\thicksim450 $  \\

\cline{1-4}
\end{tabular}
\end{table}

Based on above arguments and meanwhile in order to show the preference of the Higgs data and the fine tuning argument
on the parameter space, we classify the surviving samples into three types as follows:
\begin{itemize}
\item Type-I samples: those with $\chi^2 \leq 25$ and meanwhile $\max\{\Delta_Z, \Delta_h\} \leq 50$. This type of sample
is considered as the physical sample in our discussion.
\item Type-II samples: those with $\chi^2 \leq 25$ but $\max\{\Delta_Z, \Delta_h\} > 50$. This type
of sample can not be excluded by experiments, but is not favored by the fine tuning argument.
\item Type-III samples: those with $\chi^2 > 25$. Obviously, this type of sample is of less interest than the previous two types.
\end{itemize}
For completeness, we present in Table I the allowed ranges for these samples. As shown in Fig.\ref{fig1} and
Fig.\ref{fig2} and also in this Table, with the increase of $\lambda$ the parameter space of $\lambda$-SUSY are
crushed into a narrow region until $\lambda$ reaches its maximum, which is about 1.8.

\begin{figure}[t]
\includegraphics[height=6cm,width=14cm]{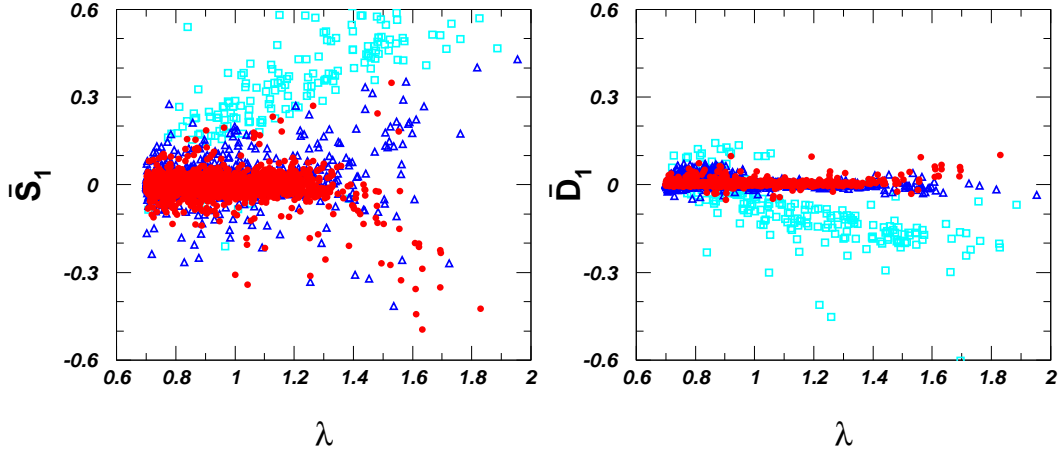}
\vspace{-0.5cm}
\caption{Singlet component coefficient $\bar{S}_1$ and non-SM doublet component coefficient $\bar{D}_1$ of the SM-like
Higgs boson  as a function of $\lambda$.  Here red bullet, blue triangle and sky-blue square
denote Type-I sample, Type-II sample and Type-II sample respectively. \label{fig3}}
\end{figure}

\begin{figure}[t]
\includegraphics[height=12cm,width=14cm]{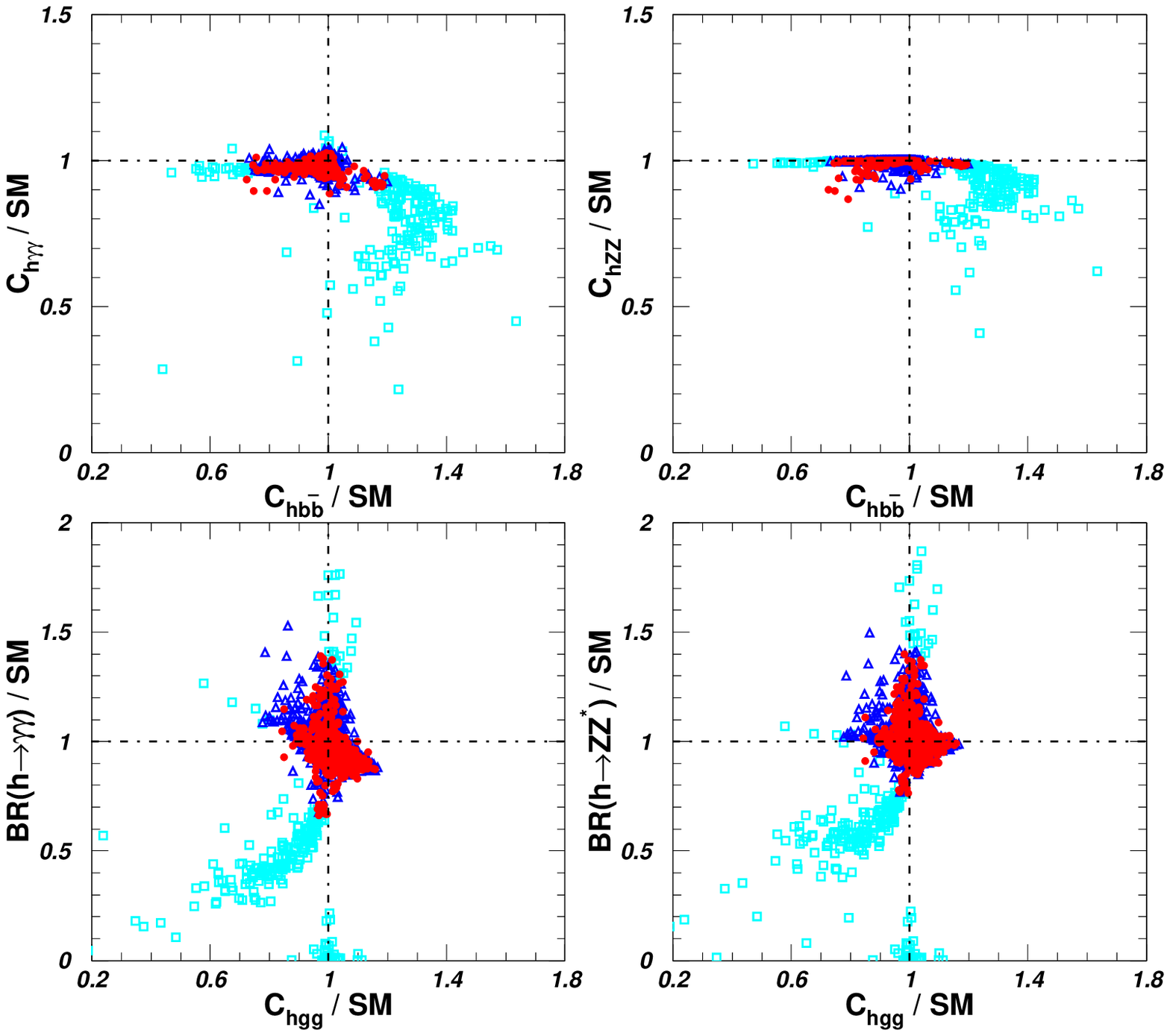}
\includegraphics[height=6cm,width=14.5cm]{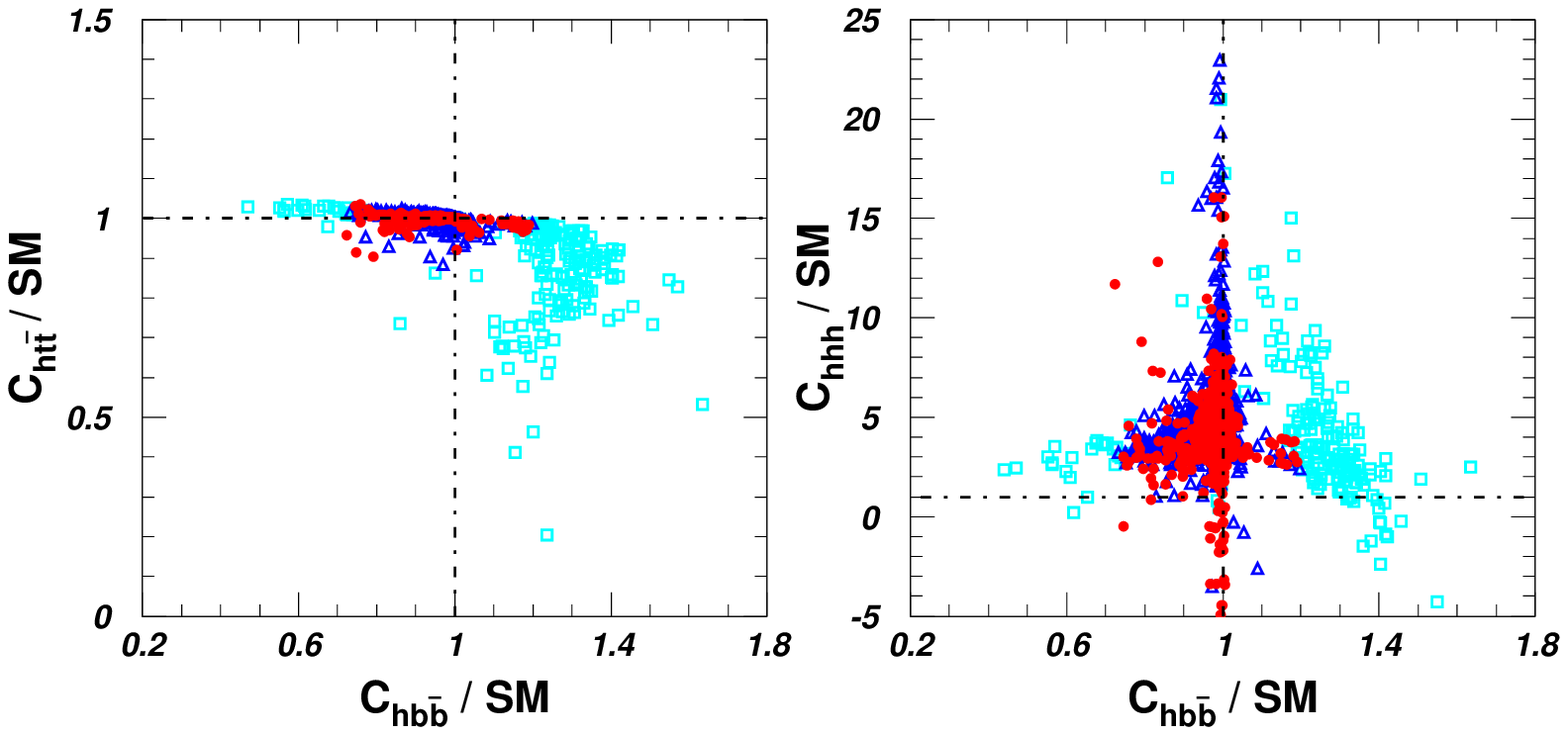}
\vspace{-0.5cm}
\caption{Coupling information of the SM-like Higgs boson for Type-I sample (red bullet),
Type-II sample (blue triangle) and Type-III sample (sky-blue square). \label{fig4}}
\end{figure}

\section{Properties of $h_1$ and $h_2$}

In this section, we explore the Higgs sector of $\lambda$-SUSY to exhibit the properties of the lightest and the next-to-lightest
CP-even Higgs bosons. We pay particular attention to the features of the bosons that differentiate $\lambda$-SUSY from
the MSSM or from the NMSSM with a low $\lambda$.

\subsection{Properties of the Lightest CP-even Higgs Boson}

As we mentioned before, throughout this work we treat the lightest CP-even Higgs boson as the SM-like Higgs boson,
so  some properties of $h$ such as its coupling to vector bosons have been limited to closely mimic those of
the SM Higgs boson. However, as we will show below, the triple self coupling of
the boson or more general the Higgs potential may still differ greatly from that of the SM.

In Fig.\ref{fig3}, we show the singlet component coefficient $\bar{S}_1$ and non-SM doublet component coefficient $\bar{D}_1$
of $h$ for Type-I samples (red bullet), Type-II samples (blue triangle)  and Type-III
samples (sky-blue square) on the left panel and right panel respectively.
This figure indicates that $|\bar{S}_1|$ may exceed $0.6$ without considering the Higgs data, and
it is up bounded by about $0.5$ at $95\%$ C.L. after considering the data. In contrast,
$|\bar{D}_1|$ reaches at most about $0.3$ and $0.1$ before and after considering the data respectively,
and given $\bar{S}_1 \neq 0$, we find it is always much smaller than $|\bar{S}_1|$ after considering the data.
The reason for the difference between $\bar{S}_1$ and $\bar{D}_1$ is
that the constraints we considered have put non-trivial requirements on the elements ${\cal{M}}_{11}^2$ and
${\cal{M}}_{12}^2$ of the CP-even Higgs mass matrix, e.g. ${\cal{M}}_{11}^2 \gtrsim 200 {\rm GeV}^2$ and
$|{\cal{M}}_{12}^2/{\cal{M}}_{11}^2 | \ll 1$, so $|\bar{D}_1|$ is forbidden to be moderately large.  In comparison,
${\cal{M}}_{33}^2$ is less constrained due to the singlet nature of the field $S_3$, and given
${\cal{M}}_{33}^2 \simeq {\cal{M}}_{22}^2 $, $|\bar{S}_1|$ may be as large as 0.7.
Furthermore, from the coupling expressions of $h$ in Eq.(\ref{couplings}) one can learn that the
$h Z Z$ coupling is always suppressed in comparison with its SM value due to the non-vanishing of
$\bar{S}_1$ and $\bar{D}_1$, while for the fermion Yukawa couplings $Y_{h\bar{f}f}$, depending the sign of $\bar{D}_1$
it may be either enhanced or suppressed. Explicitly speaking, given $\bar{S}_1 = 0$ and $ |\bar{D}_1| < 0.1 $, one can learn
that $C_{h\bar{u}u}$ is slightly enhanced while $C_{h\bar{d}d}$ is suppressed if $\bar{D}_1$ is positive, and
the situation reverses for the couplings if $\bar{D}_1$ changes its sign. In any case, the larger $|\bar{S}_1|$
becomes, the smaller the couplings are. Fig.\ref{fig3} also indicates that the values of $|\bar{S}_1|$ and $|\bar{D}_1|$ tend
to increase with the increase of $\lambda$, and so are the deviations of the normalized couplings from unity.
The reason is that  $m_{h,tree}$ will be much larger than $125 {\rm GeV}$ for a sufficient large $\lambda$,
and sizable mixings must intervene to pull down the mass.

We also compare our results in Fig.\ref{fig3} with those in \cite{lambd-SUSY-recent-2}, where a similar fit
was performed by using the Higgs data in 2013 in the framework of $\lambda$-SUSY. We find that now the allowed ranges
of $|\bar{S}_1|$ and $|\bar{D}_1|$ shrink significantly.  This reflects the more tightness of the constraints we considered
in limiting the Higgs properties.

\begin{table}
\caption{Benchmark points for different cases considered in this work. Note all the input parameters are defined at $1 {\rm TeV}$,
and in calculating the spectrum of the Higgs bosons, important radiative corrections have been taken into account.}
  \setlength{\tabcolsep}{2pt}
  \centering
  \begin{tabular}{|c|c|c|c|c|}
    \hline \hline
     No. of Point &  Point 1 (P1) &  Point 2 (P2) & Point 3 (P3) & Point 4 (P4) \\
    \hline
     $\lambda$ & 0.86 & 0.94 & 0.75 & 0.71
     \\
     $\kappa$ & 1.27 & 1.19 & 1.64 & 1.64
     \\
     $\tan\beta$ & 1.59 & 1.82 & 1.45 & 1.35
     \\
     $\mu$(GeV) & 697.6 & 1100.7 & 614.0 & 599.8
     \\
     $M_A$(GeV) & 2127.9 & 2237.9 & 388.1 & 372.6
     \\
     $M_P$(GeV) & 1449.4 & 283.8 &  2282.4 & 2135.2
     \\
     $M_1$(GeV) & 169.9 & 200.7 & 83.2 & 70.5
     \\
     $M_{Q_3}$(GeV) & 1646.3 & 675.5 & 409.2 & 1587.8
     \\
     $M_{U_3}$(GeV) & 511.0 & 1435.6 & 697.6 & 894.7
     \\
     $M_{\tilde l}$(GeV) & 186.5 & 218.5 & 128.4 & 112.4
     \\
     $A_t$(GeV)
      & -25.1 & 1539.6 & 881.1 & -1639.8
     \\
     $A_\lambda$(GeV)
     & 766.0 & -310.5 & -1248.9 & -1302.0
     \\
     $A_\kappa$(GeV)
     & -674.5 & -12.5 & -1602.3 & -1032.6
     \\
     \hline
     $m_h$(GeV)
     & 125.0 & 125.4 & 125.2 & 124.4
     \\
     $m_{H_2}$(GeV)
     & 1697.3 & 2144.6 & 232.4 & 245.0
     \\
     $m_{A_1}$(GeV)
     & 1440.5 & 174.0 & 150.5 & 102.3
     \\
     $m_{H^\pm}$(GeV)
     & 2088.2 & 2153.7 & 228.1 & 238.6
     \\
     $\bar D_1$
     & -8$\times 10^{-5}$ & 2$\times 10^{-3}$ & 0.01 & 0.04
     \\
     $\bar S_1$
     & 0.02 & -0.02 & 6$\times 10^{-3}$ & -3$\times 10^{-3}$
     \\
     $\bar D_2$
     & -0.22 & -0.98 & -0.99 & -0.99
     \\
     $\bar D_{A_1}$
     & 0.08 & 0.10 & 0.99 & 0.99
     \\
     \hline
     $\chi^2$
     & 12.2 & 12.1 & 12.0 & 13.5
     \\
     $\Delta_Z$
     & 35.4 & 139.6 & 46.9 & 45.1
     \\
     $\Delta_h$
     & 38.4 & 233.2 & 89.5 & 89.1
     \\
     $C_{hhh}/SM$
     & 13.7 & 22.1 & 5.1 & 4.2
     \\
     $\sigma(gg\to hh)/SM$
     & 34.3 & 96.1 & 1.3 & 0.6 \\
  \hline
  \end{tabular}
\label{tab1}\vspace{-0.35cm}
\end{table}

Now let's turn to the couplings of $h$. In Fig.\ref{fig4}, we exhibit such information for
same samples as those in Fig.\ref{fig3}. This figure indicates that after imposing the constraints from the Higgs data,
the normalized couplings $C_{h\gamma\gamma}/SM$, $C_{hZZ}/SM$ and $C_{h\bar{t}t}/SM$ are limited within
$15\%$ deviation from unity, and the couplings $C_{hgg}/SM$ and $C_{h\bar{b}b}/SM$ are at most $25\%$ and
$40\%$ deviating from unity respectively. Moreover, due to the
change of the width of $h$ which is mainly determined by $C_{h\bar{b}b}$, the normalized branching ratios
$Br(h\to \gamma \gamma)/SM$ and $Br(h \to Z Z^{\ast})/SM$  may vary from $0.6$ to $1.5$. Compared
with the similar fit results in 2012\cite{Early-Higgs-Fit-1}, we find that the optimal values of the couplings
are now shifted significantly.

Maybe the most impressive feature of $h$ in $\lambda$-SUSY is that the strength of its triple self coupling
$C_{hhh}/SM$ may get enhanced by a factor over 10. This is shown on the right panel of the third row in Fig.\ref{fig4},
which exhibits that $C_{hhh}/SM$ may reach 16 and 23  for the Type-I samples and Type-II samples respectively. Here we remind that
that such a great enhancement can not occur in the MSSM where the quartic terms of the Higgs potential are determined
by the weak coupling\cite{MSSM-Report}. We also remind that the enhancement seems to be limited by the naturalness argument. To see
this, we list two benchmark points with large $C_{hhh}$ in Table II (see points P1 and P2). One can easily learn
that each point corresponds to a low Higgs $\chi^2$ and meanwhile a relative large $\Delta_Z$ and $\Delta_h$, indicating that
naturalness disfavors a too large $C_{hhh}$ in $\lambda$-SUSY.

\begin{figure}[t]
\includegraphics[height=10cm,width=14cm]{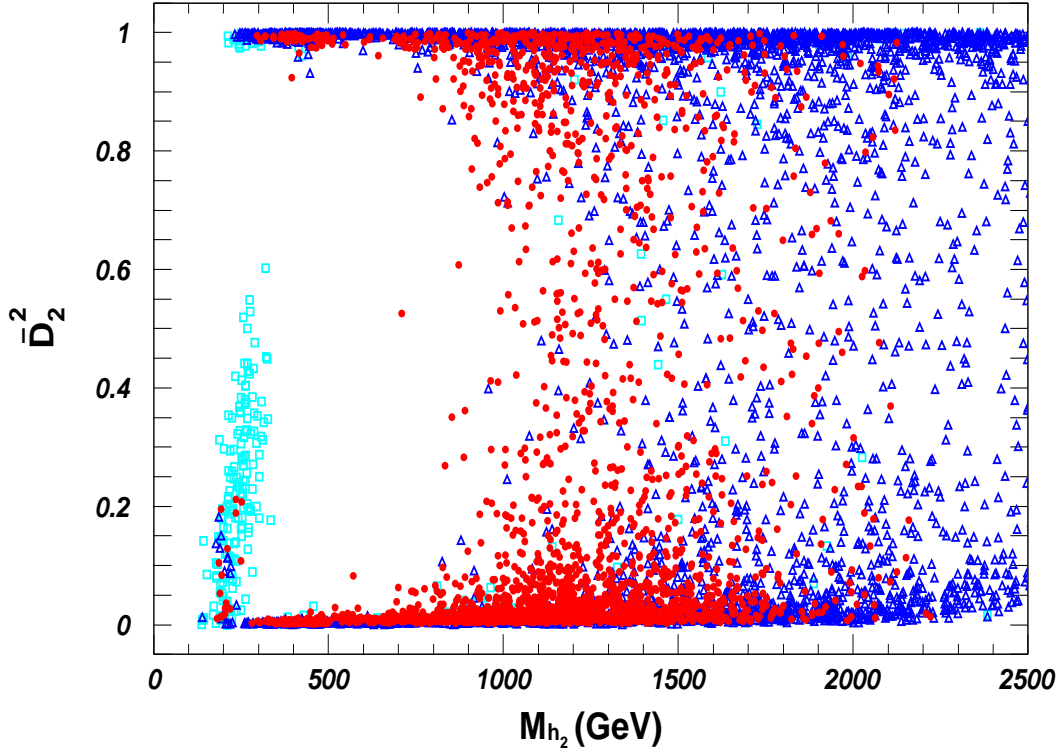}
\vspace{-0.5cm}
\caption{Doublet component of $h_2$ as a function of $m_{h_2}$ for the same samples as Fig.\ref{fig3}.
Again, Type-I, Type-II and Type-III samples are marked with red bullet, blue triangle and sky-blue
square, respectively.   \label{fig5}}
\end{figure}

\subsection{Properties of the Next-to-Lightest CP-even Higgs Boson}

Considering that $h_3$ in $\lambda$-SUSY is usually at TeV scale and thus it decouples from the electroweak physics, we here only study
the property of the Next-to-Lightest CP-even Higgs boson $h_2$. As we will show below, such a study is helpful to
understand the Higgs pair production process.

\begin{figure}[t]
\includegraphics[height=9cm,width=15cm]{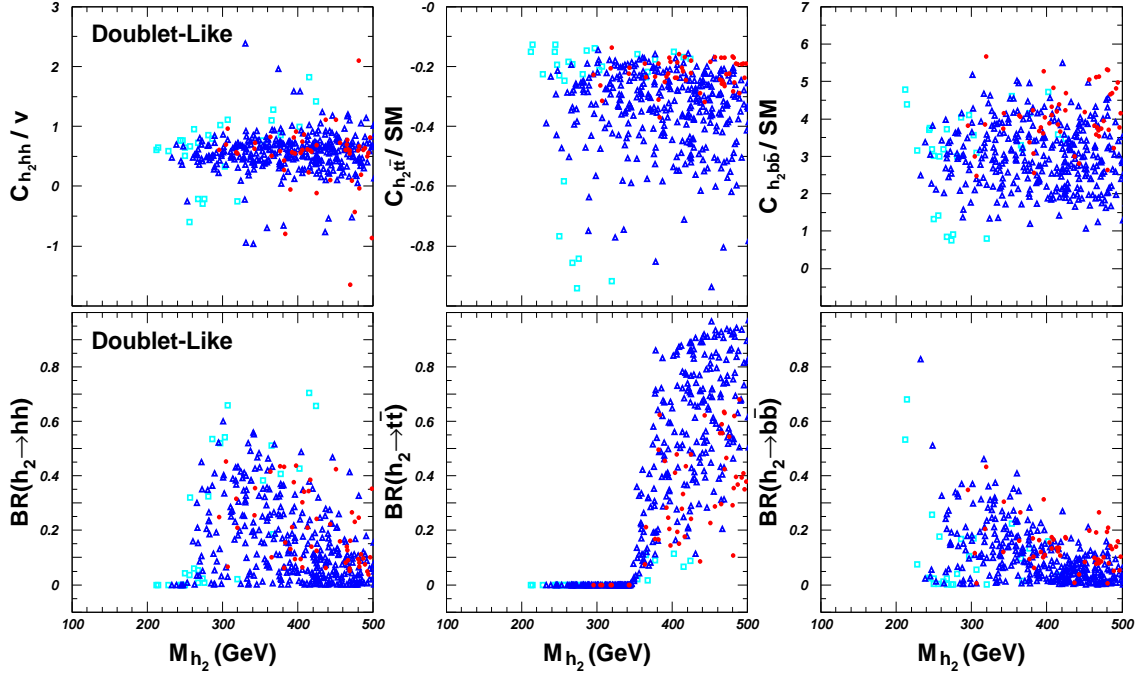}
\vspace{-0.5cm}
\caption{Couplings and Branching ratios of $h_2$ as a function of $h_2$. Note that only the samples with doublet dominated
$h_2$ in Fig.\ref{fig6} are considered. \label{fig6}}
\end{figure}

In Fig.\ref{fig5}, we show the non-SM doublet component $\bar{D}_2^2$ as a function of $m_{h_2}$ for Type-I, Type-II
and Type-III samples.  This figure reveals following information:
\begin{itemize}
\item For $m_{h_2} \leq 500 {\rm GeV}$, $\bar{D}_2^2$ is either around 1 or around 0 for most Type-I and Type-II samples.
In this case, the mixing between the fields $S_1$ and $S_3$ is small in forming $h_2$, which can be obtained if
$|{\cal{M}}_{13}^2| \ll |{\cal{M}}_{11}^2 - {\cal{M}}_{33}^2|$. Note such a situation is not altered until $m_{h_2} \gtrsim 700 {\rm GeV}$.
\item In case of $\bar{D}_2^2 \simeq 1$, $h_2$ is obviously non-SM doublet dominated, while in case of $\bar{D}_2^2 \simeq 0$
$h_2$ should be singlet dominated since $\bar{D}_3^2 = 1 - \bar{D}_1^2 - \bar{D}_2^2 \simeq 1$, which implies that $h_3$ is non-SM doublet dominated.
\item The doublet dominated $h_2$ can be as light as $200 {\rm GeV}$, which is quite different from the situation of the MSSM
where the non-SM Higgs boson $H$ must be heavier than about $300 {\rm GeV}$ after considering various constraints\cite{MSSM-Heavy-Higgs}.
As a comparison, the singlet dominated $h_2$ is more loosely limited so that it can be lighter than $150 {\rm GeV}$.
\item Note for $m_{h_2} \lesssim 300 {\rm GeV}$, there exist some type-III samples with $\bar{D}_2^2 > 0.2 $.
Since $\bar{D}_3^2 = 1 -\bar{D}_1^2 - \bar{D}_2^2 < 0.8$, these samples predict a $h_3$ with sizable singlet  and/or SM doublet components. This mixing pattern can be
achieved only for a not too heavy $h_3$. In fact, we examined the properties of these sample, and found $m_{h_3} \lesssim 650 {\rm GeV}$
and $\mu \lesssim 200 {\rm GeV}$. Since all CP-even Higgs bosons in this case are relatively light, it is apt
to be tightly limited by the Higgs data.
\item Naturalness should play a role in limiting the properties of $h_2$\cite{Fine-Tuning-H2}. Explicitly speaking, Fig.\ref{fig5} shows that
there are few Type-I samples with $m_{h_2} > 2000 {\rm GeV}$, which may be interpreted as that naturalness prefers a relatively
light $h_2$. Another example is the fraction of Type-I samples in the total number of Type-I plus Type-II samples for the
doublet dominated $h_2$ is significantly lower than that for the singlet dominated $h_2$, which means that naturalness tends
to put a tighter constraint on the doublet dominated $h_2$. All these features can be intuitively understood
by the fact that since $v_u, v_d \sim 100 {\rm GeV}$, a too heavy non-SM doublet dominated or singlet dominated
$h_2$ will make the theory fine tuned to get the correct electroweak symmetry breaking.
\end{itemize}

In the following, we try to illustrate the properties of $h_2$ for Type-I and Type-II samples with
$m_{h_2} \leq 500 {\rm GeV}$. Most of these samples are characterized by either $\bar{D}_2^2 \simeq 1$ or $\bar{D}_2^2 \simeq 0$, which is very helpful to
simplify our analysis.

We first concentrate on a doublet dominated $h_2$. Since $\bar{D}_2^2 \simeq 1$, the couplings of the $h_2$ can be approximated by:
\begin{eqnarray}
C_{h_2 V V}/SM \simeq 0, \quad C_{h_2 \bar{u} u}/SM \simeq Sign(\bar{D}_2) \cot \beta, \quad
C_{h_2 \bar{d} d}/SM \simeq - Sign(\bar{D}_2) \tan \beta. \label{coupling-doublet}
\end{eqnarray}
In Fig.\ref{fig6}, we only consider the doublet dominated $h_2$  in Fig.\ref{fig5} and show their
normalized couplings such as $C_{h_2\bar{t}t}/SM$, $C_{h_2\bar{b}b}/SM$ and $C_{h_2 h h}/v$
as functions of $m_{h_2}$. We also plot the branching ratios of $h_2 \to \bar{t} t$, $h_2 \to \bar{b} b$ and $h_2 \to h h$
in a similar way.

From Fig.\ref{fig6}, we can learn following features about Type-I and Type-II samples:
\begin{itemize}
\item In most cases, Eq.(\ref{coupling-doublet}) is a good approximation for the three $h_2$ couplings, especially
for the coupling $C_{h_2 \bar{b} b}/SM$.
\item In general, with the increase of $m_{h_2}$ the couplings $C_{h_2\bar{t}t}/SM$ and $C_{h_2\bar{b}b}/SM$ may vary within a wider ranges.
This is because the constraints we considered get relaxed as $h_2$ becomes heavy so that the couplings become more flexible to satisfy
the constraints. This character also applies  to the singlet dominated $h_2$.
\item $|C_{h_2\bar{t}t}/SM|$ is not too small: $|C_{h_2\bar{t}t}/SM| \gtrsim 0.2 $, and in optimal case, it is just
slightly below 1. On the other hand,  $|C_{h_2\bar{b}b}/SM|$ is usually larger than 1 with its maximum value reaching 6.
In this case, the $h_2 g g$ coupling is given by
\begin{eqnarray}
|\frac{C_{h_2 g g}}{C_{h g g}^{SM}}| & \simeq & \frac{\cot \beta A_{\frac{1}{2}} (\frac{m_{h_2}^2}{4 m_t^2}) - \tan \beta A_{\frac{1}{2}} (\frac{m_{h_2}^2}{4 m_t^2})}{A_{\frac{1}{2}} (\frac{m_{h}^2}{4 m_t^2})}  \nonumber \\
&\simeq & \left \{ \begin{array}{c} \{ 1.5 \cot \beta - (-0.03 + 0.03 i) \tan \beta \}/1.4 \quad \quad \quad \quad \ \ \ for \ m_{h_2} = 250 {\rm GeV}, \\
\{(2.0 + 0.01 i) \cot \beta - (- 0.02 + 0.02 i ) \tan \beta \}/1.4 \quad for \ m_{h_2} = 350 {\rm GeV},  \\
\{(2.1 + 1.1 i) \cot \beta - (- 0.01 + 0.01 i ) \tan \beta \}/1.4 \quad for \  m_{h_2} = 450 {\rm GeV},  \\
\{(1.5 + 1.6 i) \cot \beta - (- 0.01 + 0.01 i ) \tan \beta \}/1.4 \quad for \  m_{h_2} = 550 {\rm GeV},  \nonumber
\end{array} \right.
\end{eqnarray}
where the loop function $A_{\frac{1}{2}}$ is defined in \cite{MSSM-Report} and we have neglected the minor important
squark contribution. This expression indicates that due to the opposite sign of the two couplings,
the real parts of the top and bottom contributions to the  $h_2 g g$ interaction interfere constructively, while
the imaginary parts interfere destructively, and the $h_2 g g$ coupling strength is maximized at low $\tan \beta$.
\item The potentially important decay modes of $h_2$ include $h_2 \to \bar{t}t, \bar{b}b, A_1 A_1, \bar{\tilde{\chi}}_i
\tilde{\chi}_j$, where $A_1$ denotes the lighter CP-odd Higgs boson and $\tilde{\chi}_i$ represents a supersymmetric particle such as a neutralino. We find that $h_2 \to t \bar{t}$ is usually the main decay mode for $m_{h_2} \gtrsim 400 {\rm GeV}$,
and $h_2 \to h h$ (any of the decays $h_2 \to b \bar{b}, A_1 A_1, \bar{\tilde{\chi}}_i \tilde{\chi}_j$)
may be dominant over the other channels  for $260 {\rm Gev} \lesssim m_{h_2}
\lesssim 400 {\rm GeV}$ ($m_{h_2} \lesssim 250 {\rm GeV}$).
\item Considering that the case of  $m_{h_2} \lesssim 250 {\rm GeV}$ was scarcely studied before, we pay particular attention to
its features. We find that the Higgs sector in this case usually exhibits an inverted mass hierarchy, i.e. the spectrum is characterized by
$m_{h_2} \simeq m_{H^\pm} > m_{A_1}$ instead of the usual order $m_{A_1} > m_{H^\pm}$.  The underlying reason for such an anomaly
is owe to the hierarchy structure of the CP-odd Higgs mass matrix in the basis ($P_1,P_2$): $|{\cal{M}}_{P,11}^2| \ll |{\cal{M}}_{P,12}^2|
\ll |{\cal{M}}_{P,22}^2| $. For such a mass matrix, the physical scalar $A_1$ can be tuned to be very light by choosing an appropriate
value of ${\cal{M}}_{P,12}^2$. In Table II, we list two such points (denoted by P3 and P4 respectively) with P4 further
satisfying $m_{h_2} > 2 m_{A_1}$.
\end{itemize}

\begin{figure}[t]
\includegraphics[height=9cm,width=15cm]{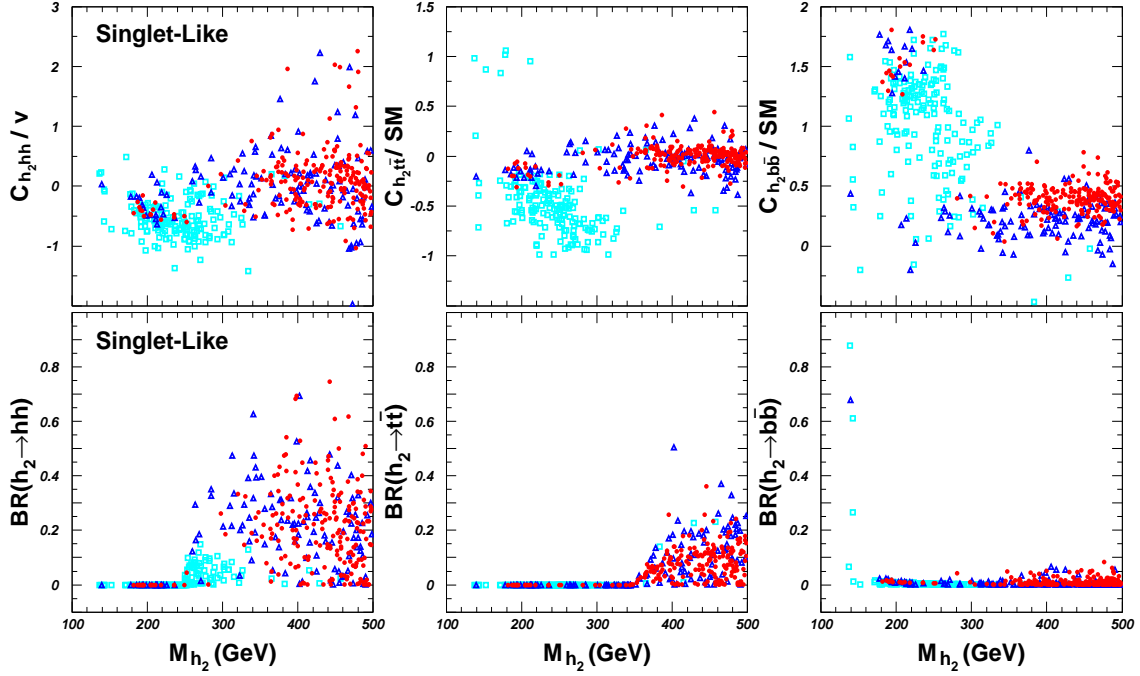}
\vspace{-0.5cm}
\caption{Same as Fig.\ref{fig6}, but for a singlet dominated $h_2$. \label{fig7} }
\end{figure}

\begin{figure}[t]
\includegraphics[height=6cm,width=14cm]{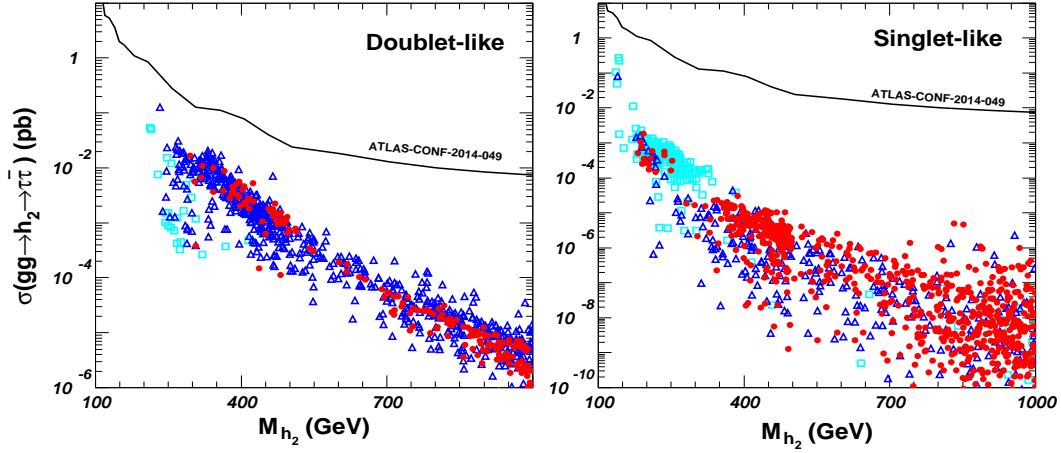}
\vspace{-0.5cm}
\caption{Same as Fig.\ref{fig6} (left panel) and Fig.\ref{fig7} (right panel), but showing the
$\bar{\tau} \tau$ signal rates induced by the process $g g \to h_2 \to \bar{\tau} \tau$ at 8-TeV LHC.
As a comparison, the bounds from the direct search for $\bar{\tau} \tau$ signal by ATLAS collaboration
are also shown.     \label{fig8}}
\end{figure}

Next we consider the singlet dominated $h_2$. In this case, since $\bar{D}_2^2 \simeq 0$ we have
\begin{eqnarray}
1 -\bar{S}_2^2 - \bar{D}_2^2 \simeq 1 - \bar{S}_2^2 \simeq \bar{S}_1^2 +\bar{S}_3^2,  \label{equation1}
\end{eqnarray}
where the sum rule $\bar{S}_2^2 = 1 - \bar{S}_1^2 - \bar{S}_3^2 $ is used. On the other hand, because
\begin{eqnarray}
\bar{S}_3^2 + \bar{D}_3^2 = \bar{S}_3^2 + 1 - \bar{D}_2^2 - \bar{D}_1^2 \simeq  \bar{S}_3^2 + 1 - \bar{D}_1^2 \leq 1,   \nonumber
\end{eqnarray}
we get
\begin{eqnarray}
\bar{S}_3^2 \lesssim \bar{D}_1^2.  \label{equation2}
\end{eqnarray}
Taking Eq.(\ref{equation1}) and Eq.(\ref{equation2}) in mind, and noticing the fact  that
$\bar{D}_1^2 \ll \bar{S}_1^2$ for a sizable $\bar{S}_1$ (see discussion about Fig.\ref{fig3}), we finally conclude that
\begin{eqnarray}
1 -\bar{S}_2^2 - \bar{D}_2^2 \simeq \bar{S}_1^2 .
\end{eqnarray}
With this approximation,  we can write down the couplings of the singlet dominated $h_2$ as:
\begin{eqnarray}
&& C_{h_2 V V}/SM \simeq |\bar{S}_1|, \quad C_{h_2 \bar{u} u}/SM \simeq Sign(V_{22}) |\bar{S}_1|,  \nonumber \\
&& C_{h_2 \bar{d} d}/SM \simeq - \bar{D}_2 \tan \beta + Sign(V_{22}) |\bar{S}_1|.
\end{eqnarray}
In Fig.\ref{fig7}, we show the couplings  $C_{h_2\bar{t}t}/SM$, $C_{h_2\bar{b}b}/SM$ and $C_{h_2 h h}/v$, and
also the branching ratios of $h_2 \to \bar{t} t$, $h_2 \to \bar{b} b$ and $h_2 \to h h$ in a way similar to Fig.\ref{fig6}.
This figure indicates that as suggested by above approximations,  both the $h_2 \bar{t} t$ and $h_2 \bar{b} b$
couplings for a singlet-like $h_2$ are usually small, but the coupling $h_2 h h$ may still be large with $C_{h_2 h h}/v$ reaching
about 2.5 in optimal case. As a result of such couplings and meanwhile the relatively strong interaction
of the $h_2 $ with sparticles\footnote{After neglecting gauge interactions, the coupling of $h_2$ with dark matter is
determined by terms $\lambda\hat{H_u} \cdot \hat{H_d} \hat{S} +\frac{1}{3}\kappa \hat{S^3}$ in the superpotential. For a singlet dominated $h_2$, the coupling strength is mainly determined by $\lambda$ for bino-like dark matter and by $\kappa$ for singlino-like dark matter. Given the potentially largeness of $\lambda$ and $\kappa$, the strength is moderately large.
While for a doublet dominated $h_2$, only its coupling with bino-like dark matter is sizable, and it is significantly smaller
than the similar coupling for a singlet dominated $h_2$ because in contrast with a sizable bino-Higgsino mixing in neutralino mass matrix,
there is no bino-singlino mixing.}, $h_2 \to \bar{t} t$ is no longer the dominant decay
channel of $h_2$ even for $m_{h_2} \gtrsim 400 {\rm GeV}$, instead any of $h_2 \to h h, A_1 A_1, \bar{\tilde{\chi}}_i \tilde{\chi}_j$
may become dominant once the kinematics is accessible. We checked that $h_2 \to \bar{\tilde{\chi}}_i
\tilde{\chi}_j $ is usually the main decay mode for $m_{h_2} < 250 {\rm GeV}$.

In order to further show the difference between a doublet dominated $h_2$ and a singlet dominated $h_2$, we plot the rate of the $\bar{\tau} \tau$
signal induced by the process $g g \to h_2 \to \bar{\tau} \tau$ at 8-TeV LHC with the same samples as those in Fig.\ref{fig6} and
Fig.\ref{fig7} respectively. For comparison, we also show the direct search bound on this signal
from the recent ATLAS analysis. This figure indicates that the $\bar{\tau} \tau$ signal rate induced by
a doublet dominated $h_2$ is usually two order larger than that by a singlet dominated $h_2$ with same mass, and in either case the rate
is at least one order lower than the direct search bound. This means that indirect experimental constraints such as
$B \to X_s \gamma$, the dark matter direct search result and the Higgs data play an important role in deciding the lower mass bound of $h_2$.

\begin{figure}[t]
\includegraphics[height=10cm,width=8cm]{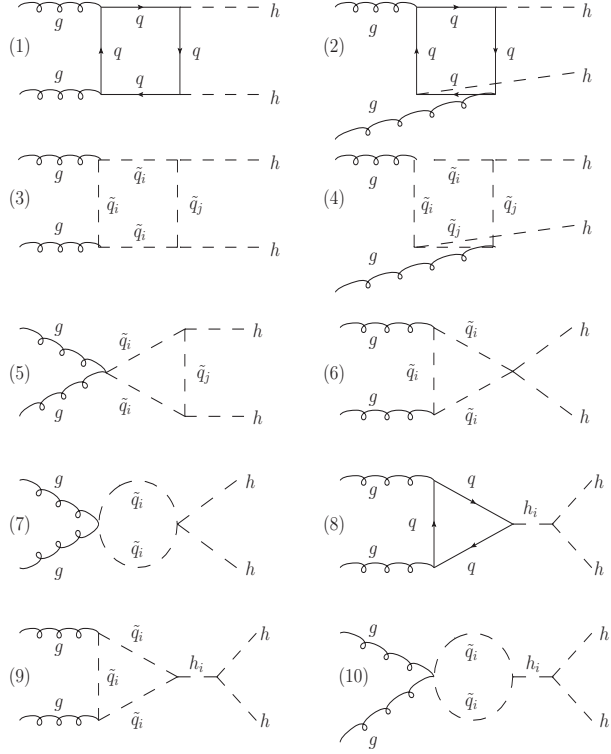}
\vspace{-0.5cm}
\caption{Feynman diagrams for the pair production of the SM-like Higgs boson
via gluon fusion in $\lambda$-SUSY with $h_i$ denoting a CP-even Higgs ($i=1,2,3$) and
$\tilde q_{i,j}$ ($i,j=1,2$) denoting a squark. The diagrams with initial gluons or
final Higgs bosons interchanged are not shown here. For the quarks and squarks we only consider
the third generation due to their large Yukawa couplings.}
\label{fig9}
\end{figure}

\section{Higgs Pair Production at the LHC}

After the discovery of the Higgs boson, the next important task of the LHC is to reconstruct the Higgs
potential and finally decipher the mechanism of the electroweak symmetry breaking. In this direction,
the Higgs pair production plays an unique role since it involves the Higgs self interactions.
So although the production is a rare process in comparison with other Higgs production processes, it
has been paid particular attention in last twenty years\cite{DHiggsInSM,SM-LO-20fb,SM-NLO-35fb}.

In the SM the Higgs pair production at the LHC proceeds by the parton process $gg \to h h$ through
the heavy quark induced box diagrams and also through the production of an off-shell Higgs which
subsequently splits into two on-shell Higgs bosons (see diagram (1), (2) and (8) of Fig.\ref{fig9}) \cite{DHiggsInSM}.
The production rate is rather low for $\sqrt{s} = 14 {\rm TeV}$, about 20 fb  at leading order~\cite{SM-LO-20fb}
and 35 fb after including the next-to-leading order QCD correction~\cite{SM-NLO-35fb}.
The capability of the LHC to detect this production process was investigated in~\cite{bbgaga,bbWW,bbtautau,DHiggs-Detect}
by the channel such as $g g \to  h h \to b \bar{b} \gamma \gamma, b \bar{b} W W^{\ast}, b \bar{b} \tau^+ \tau^-$
respectively, and it has been shown that the most efficient one is $g g \to  h h \to b \bar{b} \gamma
\gamma$ with 6 signal events over 14 background events expected for 600 fb$^{-1}$ integrated
luminosity after considering some elaborate cuts~\cite{bbgaga}. In principle, the capability
can be further improved if  the recently developed jet substructure technique is applied for
the Higgs tagging~\cite{JetSubstructure}.

In SUSY the Higgs pair production may also proceed through the diagrams 3-10 in Fig.\ref{fig9} with the internal
particles in the loops involving the third generation squarks and the intermediating s-channel scalar
being any CP-even Higgs boson \cite{Recent-SUSY-DiHiggs-Production,DiHiggs-MSSM-NMSSM}.
Since the genuine SUSY contribution to the amplitude is of the same perturbation order as the SM contribution,
the SUSY prediction on the production rate may significantly deviate from the SM result.
Based on previous studies in this field\cite{Enhanded-DiHiggs,DiHiggs-MSSM-NMSSM,Dilaton-Cao}, we learn that
there are three main mechanisms to enhance the production rate greatly:
\begin{itemize}
\item Through the loops mediated by stops\cite{DiHiggs-MSSM-NMSSM}. In SUSY, the coupling strength of
the $h \tilde{t}_i^\ast \tilde{t}_j$ interaction is mainly determined by the trilinear soft breaking parameter $A_t$,
and consequently stops contribute to the pair production in following way\cite{DiHiggs-MSSM-NMSSM}
\begin{eqnarray}
{\cal{M}} \sim \alpha_s^2 Y_t^2 ( c_1 \sin^2 2\theta_t \frac{A_t^2}{m_{\tilde{t}_1}^2} + c_2 \frac{A_t^2}{m_{\tilde{t}_2}^2}), \label{simpleform}
\end{eqnarray}
where ${\cal{M}}$ denotes the amplitude of the stop-induced box diagrams, $Y_t$ is top quark Yukawa coupling,
$\theta_t$ is the mixing angle of stops and $c_1$, $c_2$ are dimensionless coefficients determined by detailed
loop calculation. It is then obvious that the stop contributions may enhance the pair production rate greatly
for a light stop along with a large $A_t$. Detailed calculation indicates that the corrected cross section
may be several times larger than its SM prediction\cite{DiHiggs-MSSM-NMSSM}.
\item Through the resonant effect of a CP-even state $h_i$\cite{Enhanded-DiHiggs}. In SUSY, $h_i$ may be on-shell produced by $g g$ or $b \bar{b}$ initial
state. For $260 {\rm GeV} \lesssim m_{h_i} \lesssim 400 {\rm GeV}$, the production rate is not suppressed
by parton distribution function,
and meanwhile $h_i$ may decay dominantly into $hh$. In this case, the on-shell production of $h_i$ can greatly enhance the pair production rate.

In $\lambda$-SUSY, usually only the non-SM doublet dominant $h_2$ is pertinent to
the enhancement, and its resonance effect on the pair production is estimated by
\begin{eqnarray}
\sigma ( g g\to h h ) (pb) \simeq (12.5 \sim 14.5 ) \times ( \cot \beta - \tan \beta A_{\frac{1}{2}} (\tau_b)/A_{\frac{1}{2}} (\tau_t) )^2 \times
Br(h_2 \to h h),  \nonumber
\end{eqnarray}
where $\tau_b = m_{h_2}^2/(4 m_b^2)$ and $\tau_t = m_{h_2}^2/(4 m_t^2)$. In getting this estimation, we
use the fact that $\sigma(g g \to h_2)$ at 14-TeV LHC is about $14.5 {\rm pb}$ ($12.5 {\rm pb}$) for $m_{h_2} = 260 {\rm GeV}$
($350 {\rm GeV}$) given that the $h_2$ has same couplings as the SM Higgs boson to top and bottom quarks\cite{LHC-Higgs-Production-Rate},
and meanwhile neglecting the squark contribution to the $h_2 g g$ coupling.
For $\tan \beta = 2$ and $Br(h_2 \to h h) = 60\%$, one can learn that the rate is about
$(1.8 \sim 2.2 ){\rm pb}$, which is about 100 times larger than the SM prediction.
\begin{figure}[t]
\includegraphics[height=6cm,width=15cm]{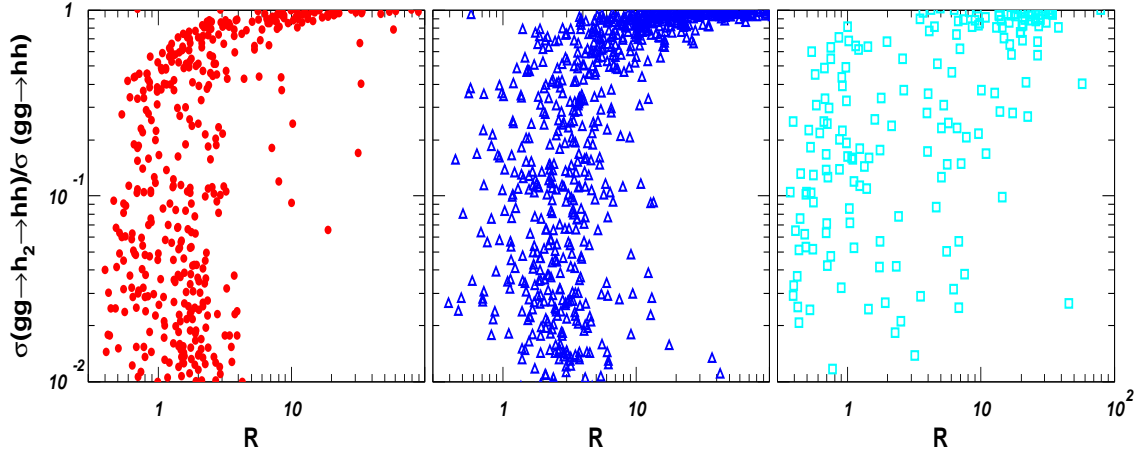}
\vspace{-0.5cm}
\caption{Correlation between the normalized total cross section $\sigma (pp \to h h)$, $R$, and the pure resonant $s$-channel contribution
to the pair production. Results shown in left panel, middle panel and right panel are for Type-I samples, Type-II samples and Type-III
samples respectively.  \label{fig10}}
\end{figure}

\item Through a large Higgs self coupling\cite{Dilaton-Cao}. In the SM, the triple self coupling of the Higgs
boson plays a minor role in contributing to
the pair production due to its relative smallness: $C_{hhh}^{SM} \simeq 32 {\rm GeV}$, and its effect is to cancel the
dominant top quark contribution. While if the self coupling is sufficiently enhanced, the situation will change and
the self coupling contribution may become dominant. Given that $h$ has same couplings as the SM Higgs boson to
top and bottom quarks, and meanwhile neglecting the squark effect to the production, one can roughly estimate
the pair production rate by\cite{Dilaton-Cao}
\begin{eqnarray}
\sigma ( g g \to h^\ast \to h h ) (pb) \simeq (C_{hhh}/SM - 2.5 )^2/1.5^2 \times 0.019,
\end{eqnarray}
where $C_{hhh}/SM$ is the normalized self coupling of the SM-like Higgs boson in $\lambda$-SUSY.
This estimation coincides in magnitude with our precise results presented in Table II for the benchmark points P1 and P2.
\end{itemize}

In the following, we define the normalized Higgs pair production rate by $R = \sigma ( p p \to h h)/ (\sigma^{LO}_{SM} ( p p \to h h )|_{m_h = 125 {\rm GeV}}) \simeq \sigma ( p p \to h h)/ (19 {\rm fb})$ for the convenience to present our results, and use the same code as
\cite{DiHiggs-MSSM-NMSSM} to calculate the cross section of $p p \to h h$ in $\lambda$-SUSY. In Fig.\ref{fig10}, we present the value
of $R$ for Type-I, Type-II and Type-III samples in left-panel, middle panel and right panel respectively.
In order to emphasize the resonance $h_2$ contribution, we also calculate the process $g g \to h_2 \to h h$ separately,
and present the ratio $\sigma( p p \to h_2 \to h h)/\sigma(pp \to h h)$ in the same figure. This figure indicates that
the Higgs pair production rate may get enhanced by more ten times
either through the resonance $h_2$ effect (corresponding to points with the ratio around one in the figure) or through the
large self coupling contribution (corresponding to points with the ratio significantly below one).
Especially, in some extreme cases we find that the pair production may get enhanced by more than 100 times for the Type-II sample,
which seems impossible in the MSSM\cite{DiHiggs-MSSM-NMSSM}.

\begin{figure}[t]
\includegraphics[height=7cm,width=9cm]{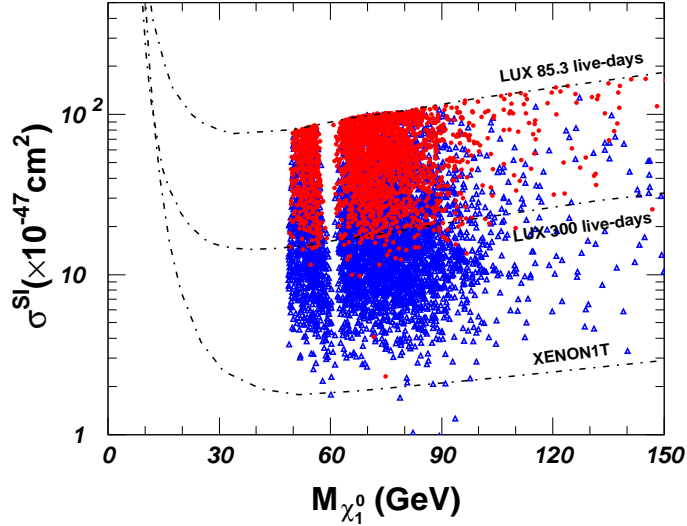}
\vspace{-0.5cm}
\caption{Same as Fig.~\ref{fig10}, but showing the spin-independent  $\chi^0_1-$nucleon scattering cross
section as a function of the dark matter only for Type-I and Type-II samples. }
\label{fig11}
\end{figure}

Before  we end our discussion, we'd like to point that the spin-independent cross section for dark matter
scattering off nucleon may be moderately large since the couplings of the CP-even states $h_i$ with dark matter
can be enhanced by a large $\lambda$\cite{NMSSM-Review}. In Fig.~\ref{fig11}, we show such a rate as a function of
the dark matter mass. In calculating the cross section, we use the formula presented in \cite{cao-dark}
by choosing a rather low $f_{Ts}$, $f_{Ts}=0.025$, which represents
the strange quark component in nucleon. This figure indicates that given LUX experiment with $300$ live-days data,
most of the Type-I samples will be excluded in case that no dark matter signal is observed. Furthermore, if
the updated XENON1T does not detect any signal of the dark matter, nearly all samples of $\lambda$-SUSY will be excluded.
These facts tell us that the dark matter direct experiments in parallel with collider experiment such as the LHC
can serve as a powerful tools in testing the framework of $\lambda$-SUSY.

\section{Summary and Conclusions}

Since the first hint of the $125 {\rm GeV}$ Higgs-like particle appeared at the end of 2011, the unnaturalness
of the MSSM in predicting the Higgs mass and also the absence of SUSY signal at the LHC have motivated more
and more interests of the non-minimal realizations of SUSY. This revived the $\lambda$-SUSY theory, which
corresponds to the NMSSM with a large $\lambda$ around one. In the framework of the $\lambda$-SUSY, the Higgs
mass can be around $125 {\rm GeV}$ even without the large top-squark radiative correction, and
meanwhile the sensitivity of the weak scale to stop masses is reduced by a factor of $(g/\lambda)^2$
in comparison with the MSSM, which means that the lower bound on the stop mass imposed by the LHC
direct searches has a weaker implication on fine-tuning in
this model than in the MSSM or the NMSSM with a low $\lambda$. Due to these advantages, the $\lambda$-SUSY has been
considered as a most natural realization of SUSY\cite{lambd-SUSY-recent,lambd-SUSY-recent-1}.

In order to implement the constraints on the $\lambda$-SUSY in a better way, we consider the Higgs data recently updated
by the ATLAS and CMS collaborations, for which the consistency of the two group results has been improved greatly.
We also define two quantities to measure the naturalness of the parameter points. After these
preparations, we scan the parameter space of the $\lambda$-SUSY by considering various constraints, then investigate
the features of its Higgs sector in physical parameter region. As is shown in this work, the improvement of the
two constraints is really necessary. For example, we find the values of the Higgs  $\chi^2$ obtained with the latest
Higgs data are significantly reduced than before, and the naturalness argument
does play an important role in selecting the parameter space of the $\lambda$-SUSY.

For the SM-like Higgs boson $h$, we have following conclusions:
\begin{itemize}
\item Current Higgs data still allow for a sizable singlet component in $h$, which at most reaches $25\%$, while the
non-SM doublet component is forbidden to be larger than $1\%$.
\item Due the latest Higgs data, the normalized couplings such as $C_{h\gamma\gamma}/SM$, $C_{hZZ}/SM$ and $C_{h\bar{t}t}/SM$
are limited within $15\%$ deviation from unity at $95\%$ C.L.. Compared with the similar fit results in 2012,
the optimal values of the couplings in the new fit are shifted significantly.
\item Interestingly, the strength of the triple self coupling of $h$ may get enhanced by a factor over 10, and naturalness
can limit such a possibility.
\end{itemize}

For the next-to-lightest CP-even Higgs boson $h_2$, we find
\begin{itemize}
\item For $m_{h_2} \leq 500 {\rm GeV}$, $h_2$ in most cases is either highly non-SM doublet dominated or highly singlet
dominated. This feature enables us to express the couplings of $h_2$ in a simple analytic way.
\item For the non-SM double dominated $h_2$, it may be as light as $200 {\rm GeV}$, which seems impossible in the MSSM.
As for its coupling, we find $|C_{h\bar{t}t}/SM| \geq 0.2$, and in optimal case the normalized coupling is just
slightly below 1. On the other hand, $|C_{h\bar{b}b}/SM|$ is usually larger than one with its maximum value
reaching 6. As a result, the $h_2 g g$ coupling may be comparable with the SM $hgg$ coupling for a low $\tan \beta$.
\item For the singlet dominated $h_2$, although it may be as light as $150 {\rm GeV}$, its couplings with SM fermions
is usually rather weak, so is of less interest in phenomenology study.
\item For either the doublet dominated $h_2$ or the singlet dominated $h_2$, the strength of the $h_2 h h$ interaction
may be quite large. Consequently, $h_2 \to h h $ can act as the dominant decay channel of $h_2$.
\item Naturalness disfavors a $h_2$ with mass at several ${\rm TeV}$ regardless its field components.
\end{itemize}

We also investigate the $h$ pair production process, and we show three mechanisms to enhance the rate greatly,
i.e. by stop-induced box diagrams, by s-channel resonant $h_2$ effect and by large self coupling of $h$. With these
mechanisms, we conclude that the $h$ pair production rate in $\lambda$-SUSY may be enhanced by more than 100 times
compared with its SM prediction.

In summary, in this work we obtained two possible characteristic features of $\lambda$-SUSY in
the experimentally allowed parameter space: 1) the triple self coupling of the SM-like Higgs
boson may get enhanced by a factor over 10 in comparison with its SM prediction; 2) the
pair production of the SM-like Higgs boson at the LHC may be two orders larger than its SM
prediction. These two features seems to be unachievable in the MSSM and in the NMSSM
with a low $\lambda$, and should be tested at the future LHC.

\section*{Acknowledgement}

We thank Ben O$^{\prime}$Leary, Florian Staub, Jinmin Yang, C.-P. Yuan, Haijing Zhou and Jingya Zhu for helpful discussions. This work was supported
by the National Natural Science Foundation of China (NNSFC) under grant No. 11222548 and 11275245.


\end{document}